\documentclass[%
 reprint,
 amsmath,amssymb,
 aps
]{revtex4-2}

\usepackage{graphicx}
\usepackage{dcolumn}
\usepackage{bm}
\usepackage{hyperref}
\usepackage{comment}
\nocite{apsrev41Control}

\graphicspath{{Pictures/}}

\begin{document}
\title{Feasibility as a moving target: Fluctuating species interactions lead to universal power law in equilibrium abundances}
\author{Cagatay Eskin}

\author{Vu Nguyen}
\author{Dervis Can Vural}%
 \email{Corresponding author: dvural@nd.edu}
\affiliation{%
 Department of Physics and Astronomy, University of Notre Dame, Notre Dame, Indiana 46556, USA\\
}%

\date{\today}

\begin{abstract}
Theoretical ecology has traditionally equated persistence with the stability of a fixed equilibrium point. Here we argue that the primary threat to ecosystem persistence need not be the loss of stability, but instead the escape of the stable equilibrium to a negative orthant. In a realistic setting, fluctuations in interactions do not merely disturb abundances about an equilibrium but can displace the equilibrium point itself. We theoretically and empirically analyze such displacements of the equilibrium point in a complex community. Theoretically, we find that light-tailed fluctuations in species interactions, no matter how small, lead to a heavy-tailed power law $P(y)=1/y^\alpha$ for the equilibrium abundance $y$ of a species. Remarkably, the exponent $\alpha=2$ is a universal value independent of interaction structure, community size, and species. Empirically, our analysis of 34 species reveals a power law signal for most, with a median exponent $\alpha \sim2.56$. Next, we derive a formula for the critical noise, $\sigma_c$, beyond which the community experiences feasibility loss ``with near certainty''. We find that $\sigma_c(N)\sim N^{-1}$, implying that larger communities are significantly more fragile to noise induced feasibility loss. Lastly, we define and calculate biologically measurable analytical metrics for both global and species-specific feasibility escape rates, and implement these metrics in dynamic simulations of 98 real world mutualistic and food web networks, to successfully predict their fragility.
\end{abstract}


\maketitle

\section{Introduction}

One of the central open questions in ecology concerns whether there is a tradeoff between the complexity and long term persistence of a community \cite{Pimm1984,Ives2007, Allesina2012}. Classic theory, following May and successors, has chosen \emph{local stability} of an equilibrium point as a key metric of persistence
 \cite{May1972,May1973,deCastro2021,Stone1990,Rao2021,ChenCohen2001,HeRuanXia1998,RoxburghWilson2000, Gellner2023,Brose2006,Grilli2016,Servan2018}. Yet stability of abundances presumes a more basic feature, \emph{feasibility}, namely that the fixed point is located in a positive orthant. If the abundances are attracted to a fixed point with  negative values, there will be extinctions and the system cannot be said to be ``stable'' in the biological sense, even if it may superficially be so in the mathematical sense.
 
Despite this foundational importance, feasibility has been studied far less than stability \cite{Dougoud2018, Saavedra2017, Grilli2017,liu_feasibility_2023}. What's known so far is that feasibility tends to vanish under ``larger'' interaction strengths  \cite{Dougoud2018} and in networks with same-sign (mutualistic and competitive) interactions \cite{liu_feasibility_2023}. Furthermore, the shape of the feasibility domain can provide insights into species-specific feasibility \cite{Grilli2017, allen-perkins_2023}. 


This study investigates the relationship between the feasibility of a community, in relation to the variability of its habitat. Environmental noise does not merely disturb abundances about a fixed equilibrium point; it alters fundamental ecological parameters that determine the position of the equilibrium point in the first place. In this case, abundances will relax perpetually towards a moving target.

We focus here on the simplest of fluctuating ecological parameters: species interactions. Interactions vary across seasons, ontogeny, behavioral contexts, productivity regimes and climate fluctuations, presumably causing changes in both the stability and feasibility of the equilibrium \cite{Deyle2016,Ushio2018,Sugihara2012,Wootton2005,Berlow2004,Murdoch1969,Oksanen1981,Werner1984,LimaDill1990,Preisser2005,Visser2005,Yang2008,Ratzke2020,Stenseth2003,Brown2004,Allesina2012, Allesina2015, Barabas2017}. Such perturbations, if large enough, might even push an equilibrium point across the feasibility boundary. How often does a community escape feasibility? How does the ``escape time'' of a community or individual species scale with the interaction structure, noise structure and the complexity of the web of interactions? For what fluctuation strength or time interval does feasibility loss become near certain? What substructures of an interaction network are most fragile?

In this paper, we address the above questions by calculating the probability distribution of the equilibrium point of a community with fluctuating interactions. Remarkably, we find that no matter how small these fluctuations are, the probability of large equilibrium shifts decays as a power law. Our prediction for single species fluctuations is $P(y)\sim1/y^\alpha$ with $\alpha=2$. 

To test our prediction, we analyze two large datasets, and detect a power law signal in abundance fluctuations, to our knowledge, for the first time in the literature. We find the corresponding exponent to mostly cluster between $\alpha=2-3$ (median $\alpha=2.56$) close to our theoretical prediction.


Next, we find that the interaction noise at which feasibility loss probability becomes near certain as it shrinks with community complexity $N$, but exponentially approaches certainty ($\sim1-1/2^N$) for higher levels interaction noise. This result establishes the feasibility counterpart to the complexity stability tension pointed out by the pioneering work of May \cite{May1972}.

Lastly, we define and calculate two metrics for risk assessment that map directly onto independently measurable quantities: (i) a species-specific ``escape rate'' that ranks which species are most likely to precipitate feasibility loss, and (ii) the mean time to feasibility loss for the community.

We should emphasize that unlike many previous theoretical studies that assume a specific statistical structure for the interaction matrix, such as gaussian/random, sparse, or having a particular correlation between interaction types \cite{May1972,May1973, Grilli2017,Dougoud2018,Saavedra2017,deCastro2021,Stone1990,Rao2021,ChenCohen2001,HeRuanXia1998,RoxburghWilson2000, Gellner2023,Brose2006,Grilli2016,Servan2018} (including those that focus on feasibility \cite{Dougoud2018, Saavedra2017, Grilli2017}) we do not make any assumptions regarding the global structure of the interaction matrix: Our results hold true for any interaction structure $A_{ij}$, each fluctuating by their own level of noise $\sigma_{ij}$.

Leaning on this strength, we validate our analytical predictions against stochastic simulations of 98 empirical ecological networks: 82 mutualistic networks and 16 food webs. Our analytical escape-rate rankings match that of these simulated for empirical networks very well, and our predicted time to feasibility loss, tracks persistence times with high concordance.

Testing our results on empirical networks reveal interesting scaling relationships linking richness, noise, and feasibility loss, offering a quantitative framework for assessing ecosystem fragility that complements simple eigenvalue-based stability analyses and directly informs empirical design and conservation prioritization.

\section{Materials and Methods}


We start with the standard generalized Lotka-Volterra (gLV) framework
\begin{align}
\frac{dn_i}{dt} = n_i\bigg[r_i + \sum_{j=1}^N A_{ij} n_j\bigg]
\label{eq: generalized_LV}
\end{align}
governing the biomass $n_i(t)$ of the $i^{\mbox{th}}$ species at time $t$, with intrinsic growth rate  $r_i$. The biomass conversion rate from species $j$ to $i$ is given by the ``interaction matrix'' $A_{ij}$. Setting the square bracket to zero yields the coexistent equilibrium condition, $\vec{n}_{\mathrm{fixed}}=\vec{x}=-{\bf A}^{-1}\vec{r}$.
The feasibility condition is $x_i>0$, for all $i$.

Even though the $A_{ij}$ are known to fluctuate \cite{Deyle2016,Ushio2018,Sugihara2012,Wootton2005,Berlow2004,Murdoch1969,Oksanen1981,Werner1984,LimaDill1990,Preisser2005,Visser2005,Yang2008,Ratzke2020,Stenseth2003,Brown2004,Allesina2012, Allesina2015, Barabas2017} it is also unrealistic to expect them to drift away indefinitely in Brownian fashion: The constancy of fundamental resources, combined with the optimized metabolic efficiency of organisms, restores interaction values to their canonical values \cite{Oksanen1981}. Furthermore, while the physical parameters influencing $A_{ij}$ such as temperature, pH, pressure, light exposure might change day to day, they will still be distributed around their ``seasonal normals''. 

As such, we will assume that $A_{ij}$ is subject to both random and restorative forces. This kind of motion is known as an Ornstein-Uhlenbeck (OU) process \cite{jacobs2010stochastic, Gardiner}, and has been used by others to describe 
fluctuations in environmental parameters \cite{Ripa_1996, Lande_2003}. Specifically, we assume that the interaction matrix ${\bf{A}}(t)={\bf{A}}+\delta {\bf{A}}(t)$, 
deviates by its mean value ${\bf{A}}$ according to the stochastic equation
\begin{equation}
\label{eq:A_SDE}
d({\delta A}_{ij}(t))=-k_{ij}\delta A_{ij}(t)dt+\sqrt{D_{ij}}dW_{ij}(t)
\end{equation}
where $k_{ij}$ and $\sqrt{D_{ij}}$ are the restorative and random influences (analogous to a spring constant and diffusion coefficient) and $dW_{ij}$ is an infinitesimal, uncorrelated, normally distributed kick \cite{Gardiner,jacobs2010stochastic}. We describe the random motion of the equilibrium point $\vec{y}(t)=\vec{x}+\delta \vec{y}(t)$ by the probability distribution $h(\vec{y})$. 

Over long times, the OU process leads to a Gaussian distribution of $\delta A_{ij}(t)$ with individual means $\langle \delta A_{ij}\rangle = 0$ and variances $\langle\delta A_{ij}^2\rangle = \sigma_{ij}^2 = D_{ij}/2k_{ij}$.

Note that that this is not equivalent to the random matrix models popular in the literature, where every element is chosen from a normal distribution. In our model the matrix can have any structure $A_{ij}$, and the normal fluctuations occur around this particular structure. Furthermore, each element has its own $\sigma_{ij}$.   


\begin{figure*}
\centering
\includegraphics[width=\linewidth]{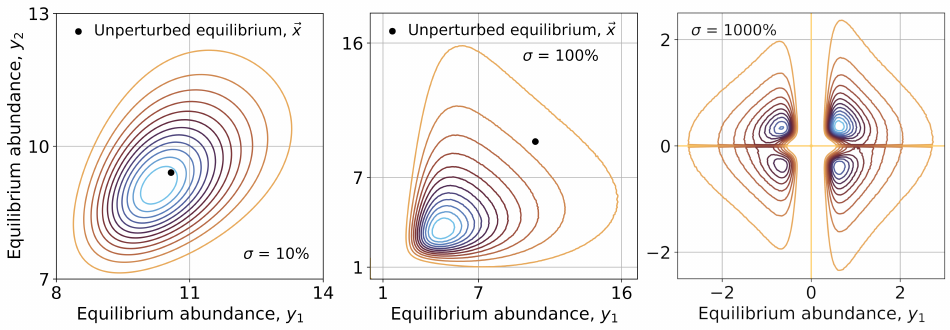}
\caption{{\bf Probability distribution of equilibrium abundances for different noise levels.} 
We plot the exact distribution $h(\vec{y})$ for ${\bf A}$=\{\{-1.4, 0.1\} , \{0.5, -1.4\} \} perturbed by $10\%$ (left), $100\%$ (center) and $1000\%$ (right) noise. As the noise increases, more probability mass moves to negative quadrants, signifying increased risk. The unperturbed equilibrium is marked by $\bm{\circ}$. 
}
\label{fig:contour_plots}
\end{figure*}

\section{Results}
\textbf{Multi-Species Equilibrium Abundance Distribution.} We obtain the probability distribution of equilibrium abundances, $h(\vec{y})$, using the equilibrium condition, $\vec{y}=-[{\bf A} + {\delta {\bf A}}]^{-1}\vec{r}$, and the distribution of $\delta \bf A$. A tedious calculation yields (see SI.1)
{\small
\begin{align}
&\!h(\vec y)
= C
\frac{e^{-\frac12\left(\kappa-\vec\psi^{T}\mathbf{\Gamma}^{-1}\vec\psi\right)}}
{\prod_i |y_i| \sqrt{{|\mathbf{\Gamma}|}}}
\Big\langle |\det[\mathbf{I}+\mathbf{\Omega}_*]|\Big\rangle
\tag{3}\\
&\!\psi_{ij}=(\beta_i F^i_{ij}-\beta_jy_iF^j_{jj}/y_j),\,\,\, \Gamma_{ij,kl}=\frac{H_{ij,kl}+H_{kl,ij}}{2}\nonumber\\
&\!H_{ij,kl}=F^k_{j l}\Big(\delta_{ik}-2\delta_{k j}\frac{y_i}{y_k}\Big)+F^j_{jj}\delta_{j l}\frac{y_i y_k}{y_j^{2}},\,\,\,F^m_{ij}=\sum_{k=1}^{N}\frac{A_{k i}A_{k j}}{\sigma_{k m}^{2}}\nonumber\\ & \kappa=\sum_{i=1}^N F^i_{ii}\beta_i^{2},\,\,\,
 C=|\det(\mathbf{A})|^N(2\pi)^{-N/2}\Big/\prod_{ij}|\sigma_{ij}|\nonumber\\
 &\!(\Omega_*)_{ij}=(1-\delta_{ij})\Omega_{ij}+\delta_{ij}\Big(\beta_i-\sum_{q\neq i}\Omega_{iq}y_q/y_i\Big), \,\,\,\beta_i=\frac{x_i-y_i}{y_i}\nonumber
\end{align}
}%
where $(i,j),(k,l)\in\Lambda=\{(u,v)\in\{1,\ldots,N\}^2, u\neq v\}$ and
the average $\langle\ldots\rangle$ is over $\bf\Omega$'s distributed (normally) as
$\propto \exp[-\frac12\sum_{(i,j),(k,l)\in\Lambda}(\Omega_{ij}-\mu_{ij})\Gamma_{ij,kl}(\Omega_{kl}-\mu_{kl})]$, with mean $\vec{\mu}=-\mathbf{\Gamma}^{-1}\vec{\psi}$.
The inverse $\mathbf{\Gamma}^{-1}$ and determinant $\det(\mathbf{\Gamma}^{-1}$) are defined as 
$\sum_{(k,l)\in\Lambda}\Gamma_{ij,kl}(\Gamma^{-1})_{kl,mn}=\delta_{im}\delta_{jn}$ and
$|\mathbf{\Gamma}^{-1}|=\det\!\big[(\mathbf{\Gamma}^{-1})_{ij,kl}\big]_{(i,j),(k,l)\in\Lambda}$.

\textbf{Universal Power Law.} The tails of $h(\vec{y})$ decay as a power law. To determine the power, we decompose the abundance vector $\vec{y} = L\hat{\mathbf{s}}$ into its magnitude $L$ and direction $\hat{\mathbf{s}}$. Analyzing the components of the joint distribution, we observe three distinct scaling behaviors as $L \to \infty$. First, the explicit prefactor scales as $(\prod |y_i|)^{-1} \sim L^{-N}$. Second, the shape parameters (matrix $\Gamma$ and exponential) depend only on equilibrium ratios and are therefore asymptotically scale-invariant ($\sim L^0$). Third, the expectation term $\langle\|I+\Omega_\ast\|\rangle$ scales as $L^{-1}$. 
Combining these terms, the joint distribution behaves asymptotically as a homogeneous function of degree $-(N+1)$, similar to a type-II multivariate Pareto distribution \cite{KotzSamuel2000Cmd},
\begin{equation}
    h(L\hat{\mathbf{s}}) \sim K(\hat{\mathbf{s}}) L^{-(N+1)}
    \label{eq:h_largeL}
\end{equation}
where $K(\hat{\mathbf{s}})$ is a direction dependent prefactor collecting all scale invariant terms.

To obtain the marginal distribution $h(y_i)$ for a single species $i$, we integrate the joint distribution over the remaining $N-1$ species. We separate the magnitude of the system and its geometry to see that integration of the volume introduces a term that scales as $y_i^{N-1}$. Including the decay of the joint probability density ($\sim y_i^{-(N+1)}$) yields the  scaling for the marginal distribution of equilibrium abundance for a single species as:
\begin{equation}
    h(y_i) \sim 1/y_i^\alpha \quad \alpha=2 .
    \label{eq:marginal_h_power_law}
\end{equation}
This power law is \emph{universal}: It holds true for any community, and any species within, independent of the specific details of the interaction structure or strength of noise (see SI.1 for details).

The heavy tails of $h(\vec{y})$ quantify the risk of catastrophic feasibility loss. A rare but large fluctuation in the interaction matrix $\delta\mathbf{A}$ can drive the system towards a configuration where $\mathbf{A}(t)$ becomes nearly singular, causing the equilibrium $\vec{y}$ to shift dramatically; and whichever $y_i$ falls outside the feasible orthant, those species $i$ are pulled towards extinction. 

For visualization purposes, we pick $N=2$, and plot $h(\vec{y})$ in Fig.\ref{fig:contour_plots} for three  levels of noise. In the low noise limit (${\bf \sigma} = 10\%$), Gaussian matrix elements lead to an approximately Gaussian $h(y)$, whose peak is approximately centered around $\vec{x}$ (Fig.\ref{fig:contour_plots}a). As the noise becomes comparable to the interaction values (${\bf \sigma} = 100\%$), the distribution becomes more lopsided and the probability mass in the negative orthants start becoming comparable to that in the positive one (Fig.\ref{fig:contour_plots}b). When the noise is much larger than the interaction values (${\bf \sigma} = 1000\%$), there is approximately equal probability mass across all $2^N$ orthants (Fig.\ref{fig:contour_plots}c).



\textbf{Low-Noise Limit.} If the noise perturbing the interactions is much smaller than the interactions themselves ($||\mathbf{A}^{-1}\mathbf{\delta A}|| \ll 1$), the probability mass under the power law tail is negligible compared to that under the bulk, and $h(\vec{y})$ can be approximated by a Gaussian, whose mean $\langle \vec{y}\rangle$ and covariance $\text{Cov}(\delta y_i, \delta y_j)$ we will identify. We first expand,
\begin{align}
\vec{y} = [\mathbf{A}+\mathbf{\delta A}]^{-1}\vec{r}= [\mathbf{I}+\mathbf{A}^{-1}\mathbf{\delta A}]^{-1}\vec{x} \simeq [\mathbf{I}-\mathbf{A}^{-1}\mathbf{\delta A}]\vec{x}\nonumber
\end{align}
The deviations from equilibrium $\delta\vec{y} = \vec{y} - \vec{x} \simeq -\mathbf{A}^{-1}\mathbf{\delta A}\vec{x}$ gives us the covariance matrix (see SI.1)
\begin{equation}
\label{eq:cov_y_gaussian}
\Sigma_{ij}=\text{Cov}(\delta y_i, \delta y_j)=\langle\delta y_i\cdot \delta y_j\rangle  = \sum_{kl} x_k^2 \sigma_{lk}^2 A^{-1}_{il} A^{-1}_{jl}.\nonumber
\end{equation} 
Up to first order in noise, there is no shift in the mean, $\langle \vec{y}\rangle=\vec{x}$. The shift emerges upon expanding up to second order in noise (see SI.1),
\begin{align}
\langle y_i\rangle =x_i+\sum_{kj} A^{-1}_{ij}A^{-1}_{kj}\sigma_{jk}^2x_k.
\label{eq: average}
\end{align}

\begin{figure}
\centering
\includegraphics[width=\linewidth]{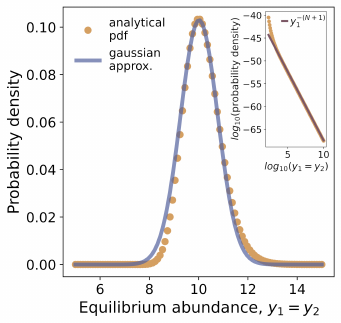}
\caption{{\bf The bulk and tail behavior of $h(\vec{y})$.} As we move along the main diagonal $|y|=y_1=y_2$ the distribution turns from a Gaussian to a power law $1/|y|^{N+1}$. The power law scaling is the same in every direction. The parameters used in the figure are same as Fig.\ref{fig:contour_plots} with $\sigma=8\%$.}
\label{fig:functional_form_fit}
\end{figure}

Figure \ref{fig:functional_form_fit} shows the equilibrium abundance distribution $h(\vec{y})$ for the $N=2$ case, plotted alongside the above low-noise (Gaussian) approximation. The inset shows the distribution's tails, which exhibit a power law scaling with an exponent of $N+1$ (further examples of tail scaling for different ecosystems are provided in SI.1). 

\begin{figure*}
\centering
\includegraphics[width=\linewidth]{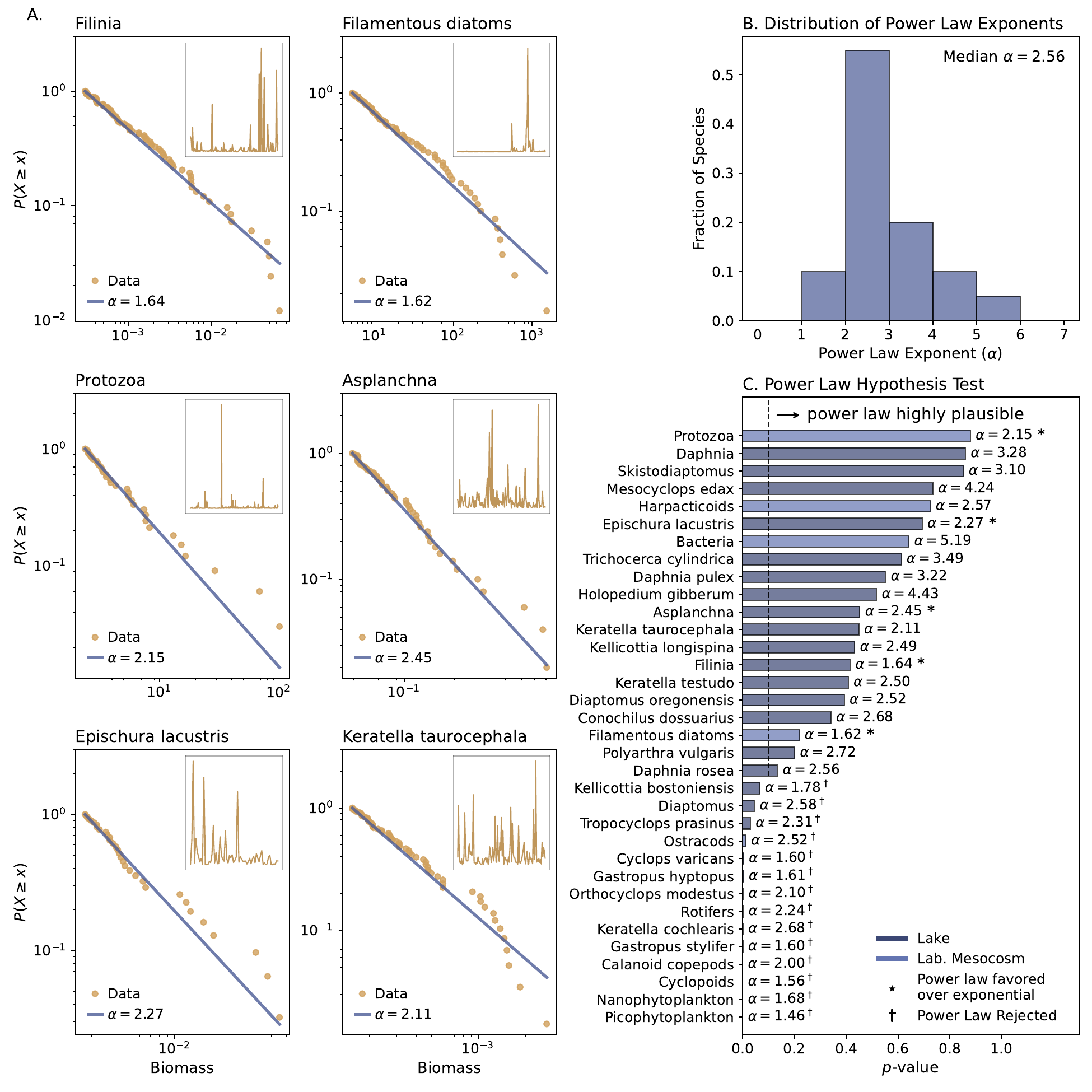}
\caption{{\bf Evidence of heavy tailed abundance distributions in aquatic communities} (A) Complementary Cumulative Distribution Functions (CCDF) for the tails of abundance on a log-log scale. Empirical data are shown as gold circles; the solid blue line represents the best fit power law distribution ($P(X \geq x) \propto x^{-(\alpha-1)}$). In the insets of each panel abundance time series of the species/functional groups are given where units are fresh weight concentration (mg/L) for Mesocosm dataset and dry mass density for Lake dataset. \cite{carpenter_synthesis_2018}.
(B) Stacked histogram showing the distribution of power law exponents ($\alpha$) for all species identified as having plausible power law tails ($p \geq 0.1$). The data combines results from a 33 year field study of a reference lake (dark blue) \cite{carpenter_synthesis_2018} and a 7 year constant condition laboratory mesocosm (light blue) \cite{beninca_chaos_2008}.
(C) Assessment of the power law hypothesis. Bars indicate the $p$-value from a Kolmogorov-Smirnov goodness of fit test using 10,000 semiparametric bootstrap simulations; values $p \geq 0.1$ indicate that the power law model cannot be rejected. Labels denote the estimated exponent ($\alpha$) for each species. An asterisk (*) indicates a statistically significant preference for the power law over an exponential distribution (Likelihood Ratio Test, $\mathcal{R} > 0, p < 0.1$). In all analyses, no statistically significant preference for the log-normal distribution over the power law was found.
}
\label{fig:tail_analysis}
\end{figure*}

\textbf{Empirical evidence for heavy tails.} We now analyze two large empirical datasets to see if abundance fluctuate according to a power law, and if so what the power $\alpha$ is. 

The first dataset is a synthesis of 33 years of whole lake measurements, where nutrient loads and food webs in several lakes were manipulated to study ecosystem responses \cite{carpenter_synthesis_2018, dornelas_biotime_2025}. For our purposes analysis, we focus specifically on Paul Lake, which served as the unmanipulated ``control group'' in \cite{carpenter_synthesis_2018}. 
The second dataset is a laboratory mesocosm containing a Baltic Sea plankton community, specifically designed to maintain strictly consistent external conditions for over 2,300 days \cite{beninca_chaos_2008}. 

We conduct our power law analyses by following the strict framework of \cite{clauset_power-law_2009}. Specifically, we first estimate the lower bound of the power law by minimizing the Kolmogorov-Smirnov (KS) statistic between the empirical data and the fitted power law model. Next, we evaluate the plausibility of the power law hypothesis using a goodness of fit test based on semi parametric bootstrapping. In this framework, the null hypothesis is that the data are drawn from a power law distribution; therefore, we reject the power law model if the resulting $p$-value is below a threshold (we pick $p \leq 0.1$, which is more cautious than the standard $p\leq0.05$). For the remaining fits that pass the test, we employ likelihood ratio tests to compare the power law model against alternative distributions, specifically the exponential and lognormal distributions. Finally, as an additional step beyond the standard procedure, we apply the Augmented Dickey-Fuller test to detect non stationarity in the abundance time series. We exclude datasets exhibiting significant directional trends to ensure that heavy tails arise from internal dynamics rather than external forcing.

Fig.\ref{fig:tail_analysis} shows the analysis of heavy tailed biomass fluctuations of species across two datasets. Fig.3A shows Complementary Cumulative Distribution Functions (CCDF), $P(X \geq x)$ and corresponding time series for six representative species. In the log-log CCDF (complementary cumulative distribution function) plots, the empirical data aligns closely with a heavy tailed a power law, and diverges significantly from an exponential tail. Fig.3B summarizes the distribution of the power law exponents, $\alpha$, across all species where a power law model was highly plausible ($p\geq0.1$). 

The tail exponents cluster between 2 and 3, with a median value of $\alpha=2.56$, not too far from our analytical value $\alpha_t = 2.0$ (Eqn. \ref{eq:marginal_h_power_law}). This cluster falls under the ``heavy tail regime'' where the distribution does not have a finite variance despite having a finite mean. 

Fig.3C shows the goodness of fit values derived from the Kolmogorov-Smirnov test, for all species from the combined datasets. The results reveal that heavy tailed distributions are a recurrent feature in these communities, irrespective of the system scale. We mark with asterisks the species for which a power law is not only plausible but statistically favored over an exponential alternative (Likelihood Ratio test, $\mathcal{R} > 0, p < 0.1$). 

The likelihood ratio tests comparing the log-normal distribution against the power law did not statistically significantly favor the log-normal model over a power law characterization of the tail. This is not surprising, since our analytical form approximates to a Gaussian near the peak and transitions into a heavy tail moving out, just like a log-normal. It is well known that distinguishing between a power law and a log-normal distribution is extremely difficult and often inconclusive unless the dataset is exceptionally large \cite{clauset_power-law_2009}. 

In SI.1, we study how a non-linear functional response affects the power law behavior. Specifically we consider a Holling type functional response $\phi_i(n_i) = \alpha_i n_i^{m_i}/({\beta_i^m + n_i^{m_i}})$, for the $i^{\mbox{th}}$ species, and find that the heavy tails persist. However, in this case, the value of $\alpha$ depends on the power of the functional response, $\alpha=m_i+1$. This dependence could account for some of the variation in Fig.3B and Fig 3C. Since a Holling exponent is typically between 1-3 for most species, this would also explain why the value of $\alpha$ is similar across so many species (Fig.2, Fig.3).

Furthermore, we show that in the low noise limit, the abundance distribution in this Holling-type response can be approximated by a log-normal distribution (SI.1). The log-normal model is marked as plausible by all 34 species we analyze, with the least plausible one having $p=0.13$, above our conservative $p=0.1$ threshold.

\textbf{Measures of Feasibility.}
Now we address (i) which species are most vulnerable to feasibility loss, and (ii) what is the persistence time of feasibility. (i) and (ii) will constitute our local and global measures of feasibility, which we will then compare with simulations of real-life communities.

In the low noise regime we can linearize the deviation from equilibrium, $\delta\vec{y}(t) \approx -{\bf{A}}^{-1}\delta {\bf{A}}(t)\vec{x}$. Evidently, $\delta \vec{y}$ is a sum of OU processes, which in general is not an OU process itself. Nevertheless, for first-passage estimates in the low-noise regime, it is convenient to approximate $\delta y_i$ by an \emph{effective} OU process that matches the stationary variance $s_i^2$ and the instantaneous diffusion $D^{\mathrm{eff}}_{ii}$ (SI.2),
\begin{align}
&d(\delta y_{i})\approx-\gamma_{i}\delta y_{i}dt+\sqrt{2D_{ii}^{\mbox{eff}}}dW_{i},\quad
\gamma_{i}=D_{ii}^{\mbox{eff}}/s_{i}^{2} \label{eq:y_SDE} 
\\
&D_{ii}^{\text{eff}} = \sum_{jk} (A^{-1}_{ij}x_k)^2 k_{jk}\sigma_{jk}^2, \quad s_i^2 = \sum_{jk} (A^{-1}_{ij}x_k)^2 \sigma_{jk}^2\nonumber
\end{align}
This approximation preserves the correct Gaussian stationary variance (since $s_i^2 = D^{\mathrm{eff}}_{ii}/\gamma_i$) and the correct short-time spreading, while compressing the multi-timescale correlation structure into a single effective relaxation rate (SI.2). The corresponding Fokker-Planck equation is \cite{Gardiner, jacobs2010stochastic}: 
\begin{equation}
\frac{\partial}{\partial t}p(\delta y_{i},t)=\frac{\partial}{\partial\delta y_{i}}[\gamma_{i}\delta y_{i}p(\delta y_{i},t)]+D_{ii}^{\text{eff}}\frac{\partial^{2}}{\partial\delta y_{i}^{2}}p(\delta y_{i},t)\nonumber
\end{equation}
Imposing an absorbing boundary at $0$ and a reflecting boundary at infinity, gives us a Kramer's formula for the extinction rate \cite{jacobs2010stochastic, Gardiner} (SI.2):
\begin{align}
\label{eq:escape_rates}
\lambda_{i}&\approx\frac{\gamma_{i} \chi_i}{\sqrt{2\pi}}e^{-\chi_i^2/2}, \qquad \chi_i=\frac{x_i}{s_i}
\end{align}
The dominant parameter in Eq.\ref{eq:escape_rates} is the barrier-to-fluctuation ratio $\chi_i$. Species with smaller abundances with larger fluctuations carry exponentially higher risk of feasibility loss. Assuming extinction events occur independently as a Poisson processes, the system-wide mean time to feasibility loss is $\langle T_{\text{loss}}\rangle \simeq (\sum \lambda_i)^{-1}$. We identify $\lambda_i$ and $\langle T_{\text{loss}}\rangle$ as a measures of individual and global robustness.


\begin{figure}[ht]
\includegraphics[width=\linewidth]{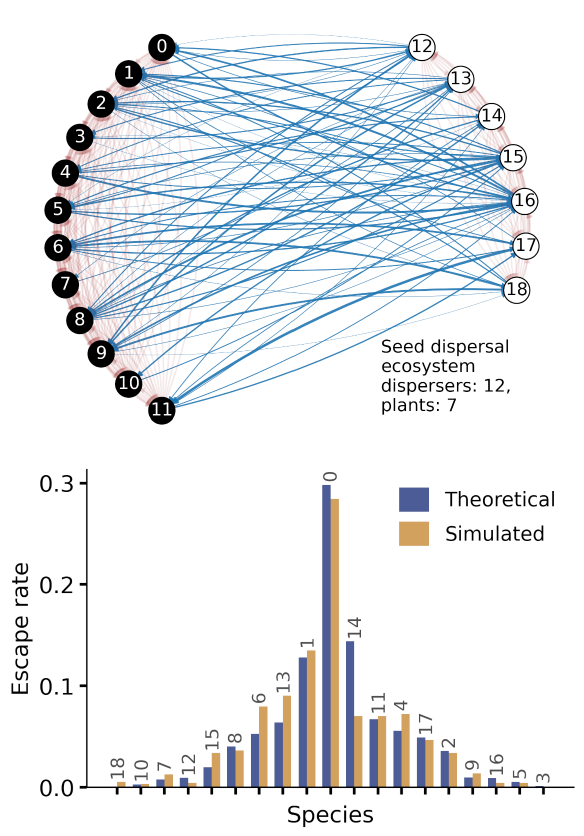}
\caption{{\bf A mutualistic seed dispersal network observed in Spain \cite{guitian1983relaciones, WebOfLifeDataset} and the risk predictions for each species}. The top panel illustrates the structure of a representative mutualistic network, composed of 12 disperser species (black nodes) and 7 plant species (white nodes). As is characteristic of this community, interactions are defined due to competition (negative, red edges) within each group and interspecific mutualism (positive, blue edges) between the two groups. The bottom panel directly compares our analytical predictions with simulation results. It shows the normalized escape rate for each species, calculated from Eq.\ref{eq:escape_rates}, alongside the corresponding feasibility loss ratios from stochastic simulations. The results demonstrate a strong correspondence, with the species predicted to have the highest escape rate also accounting for approximately one third of all simulated feasibility loss events. This predictive power, which also captures both the ranked importance and the relative contribution of each species to community fragility is a good example of predictive power of derived escape rate formula Eqn. \ref{eq:escape_rates}.}
\label{fig: network_example}
\end{figure}

\begin{figure*}
\includegraphics[width=\linewidth]{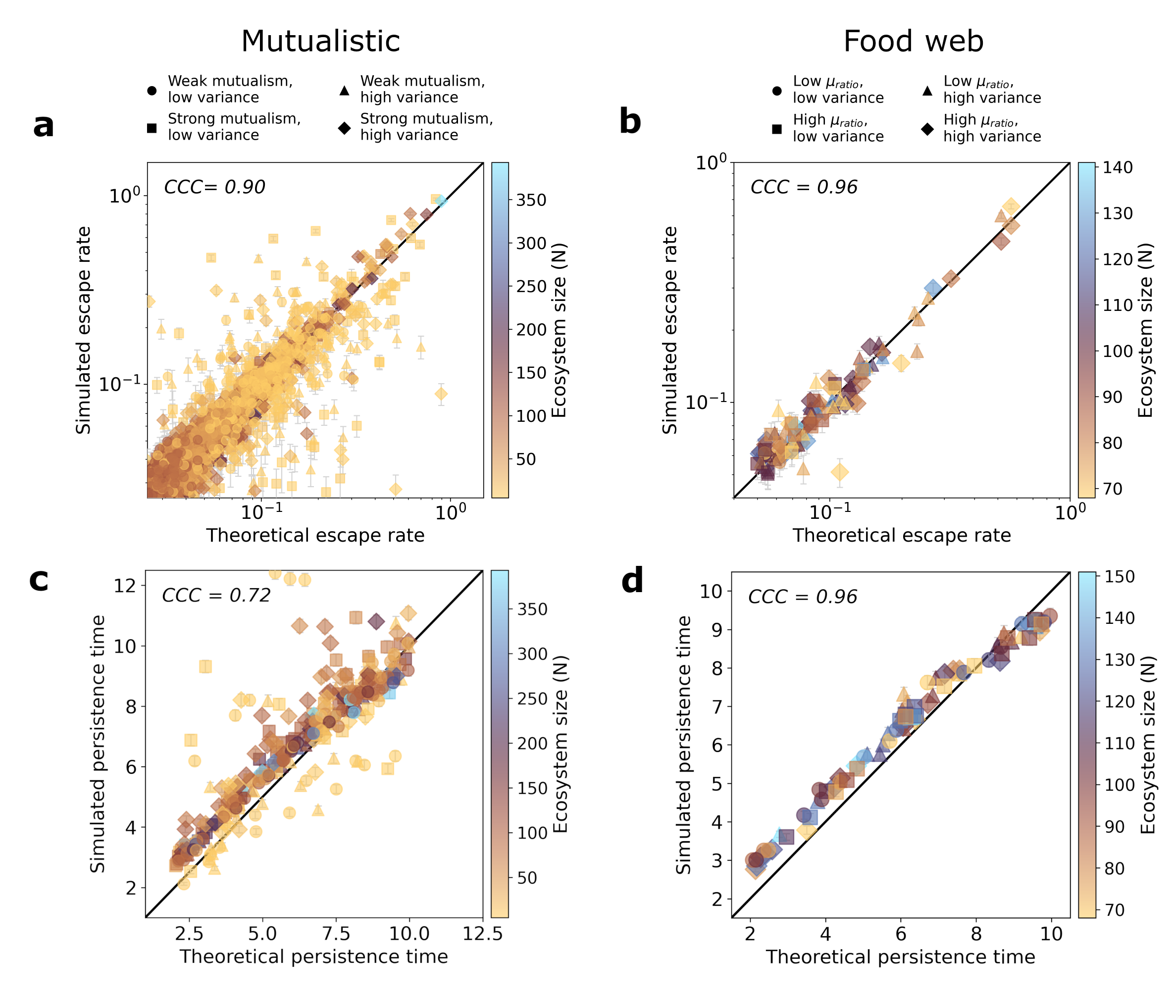}
\caption{{\bf Feasibility loss rate and mean time to loss for empirical networks alongside analytical predictions}. Each panel compares an analytical prediction (x-axis) against results from stochastic Monte Carlo simulations (y-axis), with the solid black line representing the line of perfect agreement. (a, b) The relationship between the analytical escape rate ratio (normalized Eqn. \ref{eq:escape_rates}) and the simulated corresponding loss probability. Each point corresponds to a single species within an ecosystem. Larger escape rate values indicate that the ecosystem is more likely to lose feasibility through that species, signifying its higher fragility. Panel (a) shows results for mutualistic networks, and (b) shows results for food webs. (c, d) The relationship between analytical and simulated mean time to feasibility loss. Each point represents an entire ecosystem, calculated as the inverse of the total escape rate. A larger persistence time indicates greater ecosystem robustness against fluctuations. Panel c) shows results for mutualistic networks, and d) shows results for food webs. For all panels, the color of each point corresponds to the size of the ecosystem (S), as indicated by the color bars. Marker shapes describe how the network is parameterized: for mutualistic networks, they distinguish between weak and strong mutualism and low and high variance (CV). For food webs, markers distinguish between a low and high ratio of negative to positive interaction, $\mu_{ratio}$ and low and high variance (CV). The model shows substantial agreement for food webs (Concordance Correlation Coefficient, CCC = 0.96 for both metrics), while showing moderate agreement for the mutualistic networks (CCC = 0.72 for persistence time, CCC = 0.90 for escape rates). For the panels a) and b) we excluded the species with escape rate or corresponding loss ratio values below $0.05$. For each of the parametrized networks we ensured the starting system is globally stable and feasible. For panels a) and d) weak mutualism corresponds to average positive interaction value $0.2 \times \mu_{max}$ where $\mu_{max}$ is maximum allowed average of positive interactions to ensure global stability. Similarly for strong mutualism this value is $0.6 \times \mu_{max}$. For all the panels, low variance and high variance correspond to $CV=0.5$ and $CV=2.0$, respectively. Also, negative in-group interactions representing competition are taken as $0.25$ for all the panels.} 
\label{fig: feasibility_analysis}
\end{figure*}

To test these predictions, we parameterize 98 empirical ecological networks from \cite{WebOfLifeDataset},  consisting of 82 mutualistic networks and 16 food webs. Using these empirical structures, we generate ensembles of globally stable interaction matrices for each network, similar to \cite{Grilli2017}. For mutualistic networks, we vary the mean positive interaction strength and the coefficient of variation (CV). For food webs, we vary the ratio of negative to positive interaction strengths ($\mu_{ratio}$) and the CV, generating four parameter sets per network type. We then compare our analytical predictions with stochastic simulations ran on these empirical network structures. The stochastic dynamics is out by numerically integrating Eqn.\ref{eq:A_SDE} using an Euler–Maruyama scheme, with the equilibrium recalculated at each step until feasibility loss occurred.

In the top panel of Fig.\ref{fig: network_example} we show the interaction structure of an example of mutualistic ecosystem of 12 seed dispersers and 7 plants measured in Spain \cite{WebOfLifeDataset,Fortuna2014WebOfLife}. In such a mutualistic network all intergroup interactions are sampled from a positive average that guarantees global stability of the initial system. On the contrary, intragroup interactions are sampled from a negative average value to stand for competition. Fig.\ref{fig: network_example} (bottom) shows our normalized escape rate predictions (Eqn. \ref{eq:escape_rates}) and corresponding feasibility loss rates from simulations. We find good agreement for both the rank order and numerical values of feasibility loss risk.

The broader analysis across all networks is summarized in Fig.\ref{fig: feasibility_analysis}. For food webs, the analytical model strongly predicts both escape rates and feasibility persistence times (Concordance Correlation Coefficient, CCC = 0.96 for both metrics). For mutualistic networks, agreement is also substantial (CCC = 0.72 for persistence times and CCC = 0.90 for escape rates). 
\subsection*{Conditions for Almost Certain Feasibility Loss}
In the high noise regime ($\delta A_{ij}\gg A_{ij}$), significant probability mass leaks out of the all positive orthant as $h(\vec{y})$ uniformizes across all $2^N$ orthants, therefore the probability of feasibility loss approaches $1 - 1/2^N$ (Fig.6), and accordingly, the mean time to feasibility loss approaches $0$.
\begin{figure}
\centering
\includegraphics[width=\linewidth]{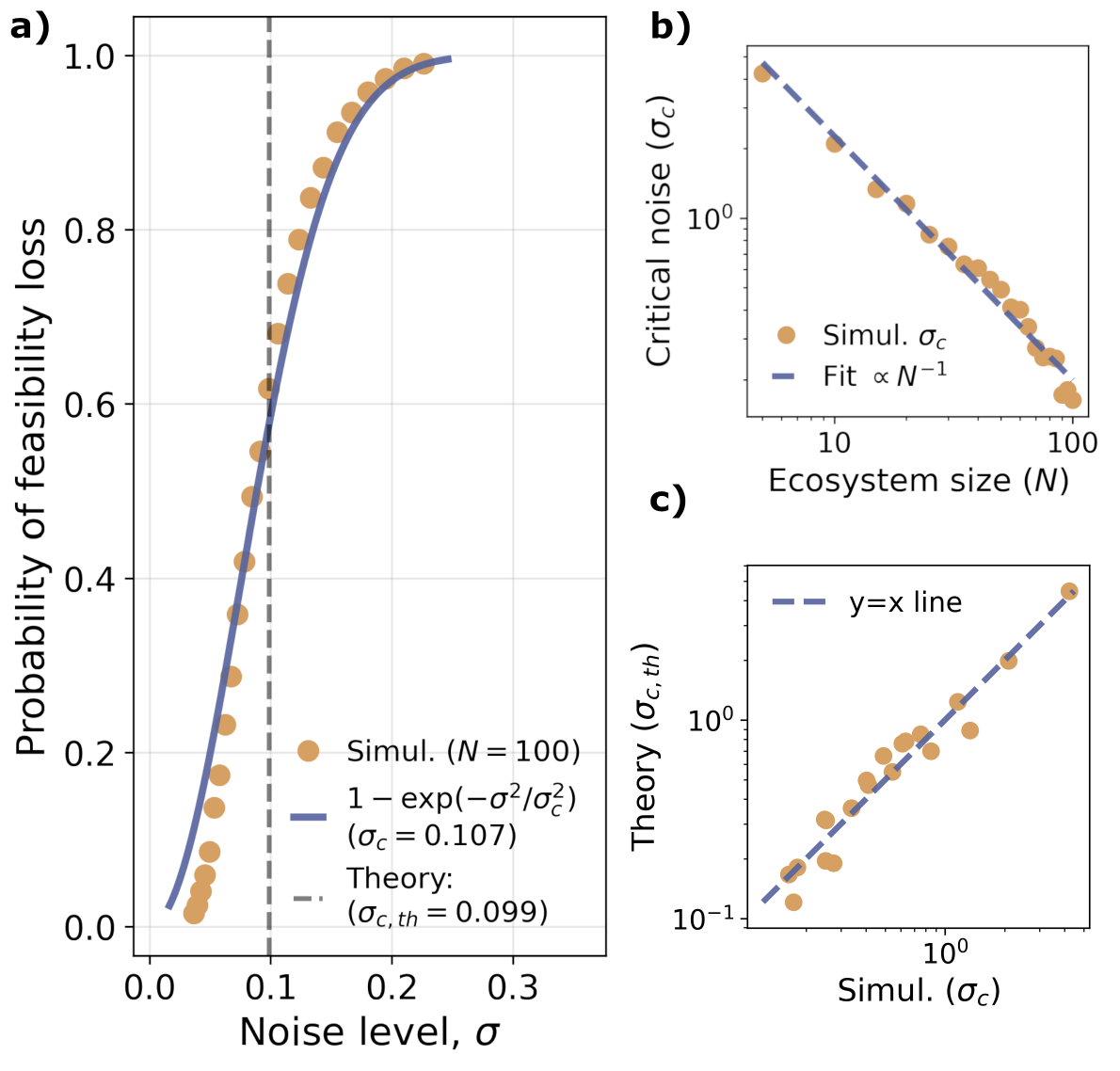}
\caption{{\bf Analysis of probability of feasibility loss for random networks}. We consider May type fully connected interaction matrices with Gaussian distributed elements to simulate the change of probability of feasibility loss with the increasing noise variance. We observe that the loss probability fits an exponential saturation model with a decay rate $\sigma^2_c$ after which feasibility loss is almost certain. Repeating the simulations for different ecosystem sizes we observe that the critical noise, $\sigma_c$ scales with ecosystem size as a power law with exponent $-1$. Panel a) shows the probability of feasibility loss obtained through Monte Carlo simulations as a function of the noise level for a $N=100$ system. The panel b) shows the result of simulations for different ecosystems with increasing number of species and their respective $\sigma_c$ fit values. We observe that as the ecosystem size increases, critical noise level which leads to almost certain feasibility loss decays with the power law $\propto N^{-1}$. Panel c) shows the difference between $\sigma_c$ obtained by fitting the simulation results versus theoretical prediction Eqn. \ref{eq: expected_var_feas_loss}.}
\label{fig:P_loss_analysis}
\end{figure}
In Fig.\ref{fig:P_loss_analysis} we numerically simulate $\vec{y}(t) = -[\mathbf{A} + \delta\mathbf{A}(t)]^{-1}\vec{r}$ to show how quickly feasibility loss converges to these asymptotes with increased noise ($\sigma^2=D/2k$) for various community sizes $N$. Feasibility loss probability is best fit by exponential saturation (Fig.6 left). Specifically, as noise is increased, the system converges to ``certain infeasibility'' at a rate of $\sigma^2_c$. We find that larger communities are more sensitive to noise; specifically, $\sigma_c\sim 1/N$. As $N\to\infty$, an infinitesimal noise $\sigma\to0$ leads to feasibility loss with near certainty.

In the opposite limit of low noise ($\delta A_{ij}\ll A_{ij}$), feasibility can still be lost with very high probability. We have shown earlier that in the low noise limit $h(\vec{y})$ is approximately a Gaussian, peaked around the shifted equilibrium with components $\langle y_i\rangle =x_i+\sum_{kj} A^{-1}_{ij}A^{-1}_{kj}\sigma_{jk}^2x_k$ where $\vec{x}$ is the unperturbed equilibrium. With increasing noise, the distribution broadens significantly faster than the mean shifts. A rough estimate for what amount of noise will cause almost certain feasibility loss can be obtained considering the broadening of the distribution and setting $\text{Var}(y_i)=\langle y_i\rangle ^2$, and $\sigma_i\sim\sigma$ in Eqn.\ref{eq:cov_y_gaussian},
\begin{equation}
\label{eq: expected_var_feas_loss}
\sigma_{\text{c,th}} = \frac{\langle y_i\rangle}{\sqrt{\sum_{kl} (A^{-1}_{il})^2 \langle y_k\rangle^2}}
\end{equation}
This estimate is marked in Fig.6, Panel a) with the dashed line and agrees well with the value $\sigma_{\text{c}}$ marking the transition to certain feasibility loss. Panel b) shows the scaling of critical variance with increasing ecosystem size obtained through the simulations and the Panel c) shows difference between $\sigma_{\text{c}}$ obtained from the simulations and theoretical prediction, $\sigma_{\text{c,th}}$.

\section{Discussion}

Our results present a feasibility centered perspective on the complexity--persistence debate. When interactions fluctuate, the equilibrium is no longer a fixed point but a stochastic, moving target; consequently, persistence can fail
even when the instantaneous dynamics remain stable. 

We have come to the startling conclusion that light-tailed (Gaussian) perturbations to interaction values, no matter how small, will lead to heavy tailed equilibrium distributions. Moreover, this result is quantitatively universal: the power law $h(y)\sim1/y^2$ does not depend on community size and interaction structure.

To quantify the strength of feasibility, we introduced two biologically interpretable, empirically measurable metrics (cf. Eqn.\ref{eq:escape_rates} and paragraph below). These metrics translate interaction statistics into a ranking of species vulnerability and global feasibility.


Our analysis of empirical data reveals a recurrent power law scaling with an exponent mostly clustered within $\alpha\sim2-3$ (median $\alpha=2.56$), placing real world ecosystems within a heavy-tailed regime. Empirical estimates are consistent with the theoretical scaling prediction $\alpha=2$. However, this raises the question of why the theory systematically underestimates the exponent, and why a finite variance in $\alpha$ across species exist in the first place, despite our finding that $\alpha$ is not structure, size, and species dependent.

First, our analysis in S.I. shows that modifying the simple Lotka-Volterra functional response to a Holling-Type functional response, turns the $h(y)\sim1/y^2$ law into $h(y)\sim1/y^{m+1}$, where $m$ is the Holling exponent. This explains not only the finite species to species variability in Fig.3, but also why the variability remains limited: In nature, $m$ tends to range between 1-3 (corresponding to an apparent $\alpha=2-4$), successfully accounting for most of the variability in our $\alpha$ data. 

Second, we should remind and caution that we use empirical abundance values $n(t)$ as a proxy for equilibrium values $y(t)$. This is fair for species whose relaxation times are faster than the random motions of interactions. But slow-relaxing species will never have time to catch up with the equilibrium, on the rare occasion that it jumps to some far extreme. This effectively shaves off the extremes, increasing the apparent value of $\alpha$. If the effect is sufficiently large, it could even wash away the power law signal below our significance threshold. The variation in relaxation times of different species then, would cause a spread in $\alpha$ values, while also pushing them ahead.

Our findings have critical implications for how we understand ecological risk and connects directly to the literature on the theory of community persistence. We suggest that the primary threat to a community's persistence may not be the failure of local stability, but rather a sudden, fluctuation driven shift of the entire equilibrium into an unfeasible state. This might provide a mechanism for the black swan events where a community that appears robust and stable for long periods can, nevertheless, suddenly crash \cite{anderson_2017}. As such, merely observing the stability of population dynamics around a mean value seems insufficient. One must account for the inherent risk of large, intermittent shifts in the equilibrium value itself. 

Feasibility loss constitutes an additional fragility channel to May's eigenvalue stability: The larger a community, the more easily its equilibrium point can be pushed into infeasibility, even though the equilibrium point itself can maintain its stability at all times.

\section*{Data Availability}
Time series abundance data for the whole lake experiments were acquired from the BioTIME database (dataset 638) and \cite{carpenter_synthesis_2018, dornelas_biotime_2025}. Data for the laboratory mesocosm experiment were obtained from \cite{beninca_chaos_2008}. The interaction networks used for the feasibility analysis of 98 empirical ecosystems were retrieved from the Web of Life repository \cite{WebOfLifeDataset,Fortuna2014WebOfLife}. The source codes necessary to reproduce all the results are available in the GitHub repository \href{https://github.com/cagatayeskin/Feasibility_analysis}{github.com/cagatayeskin/Feasibility\_analysis}.

\section*{Acknowledgments}
We thank Notre Dame Center for Research Computing for allowing us to utilize their computational sources during the production of this paper.

\bibliographystyle{apsrev4-1}
\bibliography{mainbib.bib}


\end{document}


\title{Feasibility as a moving target: Fluctuating species interactions lead to universal power law in equilibrium abundances \\
--Supplementary Material 1-- \\}

\author{Cagatay Eskin}
\email{ceskin@nd.edu}

\author{Vu Nguyen}
\email{}
\author{Dervis Can Vural}%
 \email{Corresponding author: dvural@nd.edu}
\affiliation{%
 Department of Physics and Astronomy, University of Notre Dame, Notre Dame, Indiana 46556, USA\\
}%

\date{\today}

\maketitle

\section{Calculation of probability distribution of equilibrium $\vec{y}$ for generalized lotka-volterra system}

Quantitatively, if the interactions coefficients $A_{ij}$ have some error margin described by the standard deviation of a normal distribution, $\sigma_{ij}$, then what is the probability distribution function (PDF) of equilibrium abundances $\vec{x}$? Specifically, we let $\vec{y}=-[{\bf A} + {\bf W}]^{-1}\vec{r}$ where the elements of the noise matrix $W_{ij}$ have zero mean and standard deviations $\sigma_{ij}$. The question now can be stated as what is the distribution for $\vec{y}$?

This can be done by first finding the distribution for ${\bf B} = {\bf A^{-1}W}$, and then applying the substitution $\vec{y}=[{\bf I}+{\bf B}]\vec{x}$ to get an integral for the PDF of $\vec{y}$. Finally, the integral is approximated using the saddle point method to get an analytical form for the PDF when the noise is small.

\subsection{Distribution of ${\bf B}={\bf A^{-1}W}$}
Observe that the noisy equilibrium can be written as

\begin{equation} \label{eq: stochastic_sim_equation}
    \vec{y} = -[{\bf I + A^{-1}W}]^{-1}{\bf A^{-1}}\vec{r} = [{\bf I + A^{-1}W}]^{-1}\vec{x}
\end{equation}

where ${\bf I}$ is the identity matrix. The displacement from $\vec{x}$ is therefore generated by the random elements in ${\bf B}$. If the random elements ${\bf W}$ are independent with probability density function (PDF) given by
\begin{equation*}
    p({\bf W},\sigma) = \prod_{ij}\frac{\exp\left(\frac{-W_{ij}^2}{2\sigma_{ij}^2}\right)}{\sqrt{2\pi\sigma_{ij}^2}}
\end{equation*}
then the transformation to ${\bf B}$ is given by
\begin{equation*}
g({\bf B},\sigma) = p({\bf AB},t) \left| \frac{\partial{\bf W}}{\partial{\bf B}} \right|.
\end{equation*}

The Jacobian of this transformation has a block diagonal structure. First, observe that a derivative of $W_{ij} = \sum_k A_{ik}B_{kj}$ with respect to $B_{pq}$ is $\frac{\partial W_{ij}}{\partial B_{pq}}=A_{ip}\delta_{jq}$ where $\delta_{jq}$ is the Kronecker delta function. Then the Jacobian matrix can be ordered into block diagonal form
\begin{equation*}
{\bf J} =\begin{bmatrix}
	\frac{\partial W_{11}}{\partial B_{11}} 	&	\frac{\partial W_{11}}{\partial B_{21}}	& \dots & \frac{\partial W_{11}}{\partial B_{N1}} & \frac{\partial W_{11}}{\partial B_{12}}	& 	\frac{\partial W_{11}}{\partial B_{22}}		& \dots &	\frac{\partial W_{11}}{\partial B_{NN}} \\
	
	\frac{\partial W_{21}}{\partial B_{11}} 	&	\frac{\partial W_{21}}{\partial B_{21}}	& \dots & \frac{\partial W_{21}}{\partial B_{N1}} & \frac{\partial W_{21}}{\partial B_{12}}	& 	\frac{\partial W_{21}}{\partial B_{22}}		& \dots &	\frac{\partial W_{21}}{\partial B_{NN}} \\
	
		\vdots & \vdots & \ddots & \vdots & \vdots & \vdots &\ddots &\vdots \\
		
	\frac{\partial W_{N1}}{\partial B_{11}} 	&	\frac{\partial W_{N1}}{\partial B_{21}}	& \dots & \frac{\partial W_{N1}}{\partial B_{N1}} & \frac{\partial W_{N1}}{\partial B_{12}}	& 	\frac{\partial W_{N1}}{\partial B_{22}}		& \dots &	\frac{\partial W_{N1}}{\partial B_{NN}} \\
	
	\frac{\partial W_{12}}{\partial B_{11}} 	&	\frac{\partial W_{12}}{\partial B_{21}}	& \dots & \frac{\partial W_{12}}{\partial B_{N1}} & \frac{\partial W_{12}}{\partial B_{12}}	& 	\frac{\partial W_{12}}{\partial B_{22}}		& \dots &	\frac{\partial W_{12}}{\partial B_{NN}} \\
	
	\frac{\partial W_{22}}{\partial B_{11}} 	&	\frac{\partial W_{22}}{\partial B_{21}}	& \dots & \frac{\partial W_{22}}{\partial B_{N1}} & \frac{\partial W_{22}}{\partial B_{12}}	& 	\frac{\partial W_{22}}{\partial B_{22}}		& \dots &	\frac{\partial W_{22}}{\partial B_{NN}} \\
	
		\vdots & \vdots & \ddots & \vdots & \vdots & \vdots &\ddots &\vdots \\
		
\frac{\partial W_{NN}}{\partial B_{11}} 	&	\frac{\partial W_{NN}}{\partial B_{21}}	& \dots & \frac{\partial W_{NN}}{\partial B_{N1}} & \frac{\partial W_{NN}}{\partial B_{12}}	& 	\frac{\partial W_{NN}}{\partial B_{22}}		& \dots &	\frac{\partial W_{NN}}{\partial B_{NN}} \\
\end{bmatrix}.
\end{equation*}
This arranges the $N^2\times N^2$ Jacobian into $N$ diagonal blocks of ${\bf A}$ and zero every where else. Therefore, the positive determinant required for the transformed PDF is $|\partial {\bf W}/\partial {\bf B}|=||{\bf A}||^N$. Hence, this transformation is possible if the ${\bf A}$ is invertible. Which is the same condition as the existence of $\vec{x}$.

The linear transformation from Gaussian $W_{ij}$ will in turn give a Gaussian distribution for $B_{ij}$. The arguments inside the exponential can be directly inferred using the mean and covariance of $B_{ij}$. Let ${\bf R} = {\bf A^{-1}}$ represent the inverse of the mean interactions. Then the mean for each $B_{ij}$ is given by
\begin{equation*}
    \langle B_{ij}\rangle = \left\langle \sum_k R_{ik}W_{kj} \right\rangle = \sum_k R_{ik} \left\langle W_{kj}\right\rangle=0
\end{equation*}
and therefore the covariance is
\begin{align*}
    Cov( B_{ij} B_{pq} ) &= \left\langle \left(\sum_k R_{ik}W_{kj}\right)\left(\sum_m R_{pm}W_{mq}\right) \right\rangle\\
    &=\sum_k R_{ik}R_{pk}\sigma^2_{kj}\delta_{jq}
\end{align*}
where each $W_{ij}$ are independently distributed. The covariance matrix will also be a block diagonal matrix due $\delta_{jq}$. Using the same ordering as the Jacobian with
\begin{equation*}
\vec{v} = \begin{bmatrix}
B_{11} \\
B_{21} \\
\vdots \\
B_{N1}\\
B_{12}\\
B_{22}\\
\vdots\\
B_{NN}
\end{bmatrix} = \begin{bmatrix}
\vec{v}^{(1)} \\
\vec{v}^{(2)} \\
\vdots \\
\vec{v}^{(N)}\\
\end{bmatrix}
\end{equation*}
the next step requires an exponential argument of the form $\vec{v}^T {\bf C} \vec{v}$ where ${\bf C}$ is the inverse of the covariance matrix for the ${\bf B}$ PDF.

The $m$-th block is given by
\begin{equation*}
    \vec{v}^{(m)} = \begin{bmatrix}
B_{1m} \\
B_{2m} \\
\vdots \\
B_{Nm}\\
\end{bmatrix}
\end{equation*}
and the elements of the covariance block ${\bf T}^{(m,m)}$ are given by
\begin{equation*}
    T^{(m,m)}_{ij} = Cov(B_{im},B_{jm}) = \sum_k R_{ik}R^T_{kj}\sigma_{km}^2.
\end{equation*}
When $\sigma_{km}=1$ this covariance matrix simplifies to ${\bf T}^{(m,m)}={\bf RR^T}$. Therefore, the corresponding precision block is ${\bf C}^{(m,m)}={\bf A^TA}$ where by definition ${\bf AR} = {\bf RA} = {\bf I}$. When $\sigma_{km}\neq 1$ the corresponding precision block is given by
\begin{equation*}
    C^{(m,m)}_{ij} = \sum_k \frac{A^T_{ik}A_{kj}}{\sigma^2_{km}}.
\end{equation*}
This has been verified by checking ${\bf TC}={\bf CT}={\bf I}$ in Appendix \ref{app:TC}.

With these two expressions the PDF for ${\bf B}$ is given by
\begin{equation}
    g({\bf B},\sigma) = \frac{\left|\left|{\bf A}\right|\right|^N}{\prod_{ij}| \sigma_{ij}|}  (2\pi)^{\frac{-N^2}{2}} \exp(\frac{-\vec{v}^T {\bf C} \vec{v}}{2}) \label{eq:B_pdf}
\end{equation}
where the block diagonal precision matrix (of size $N^2\times N^2$) is given by
\begin{equation}
    {\bf C}^{(m,q)} = {\bf F}^{(m)} \delta_{mq} \label{eq:C_definition}
\end{equation}
for the block located at the $m$-th row and $q$-th column. Each diagonal block (of size $N\times N$) is given by
\begin{equation}
    F^{(m)}_{ij} = \sum_k \frac{A^T_{ik}A_{kj}}{\sigma^2_{km}}. \label{eq:F_definition}
\end{equation}

\subsection{Integral form of the equilibrium distribution}

The PDF of $\vec{y}$ can be written as
\begin{equation*}
h_{{\bf A},\vec{x}}(\vec{y},\sigma) = \int \delta(\vec{y}-[{\bf I + B}]^{-1}\vec{x}) g({\bf B},\sigma) d{\bf B}.
\end{equation*}
This is an integral over $N^2$ values of ${\bf B}$ with $N$ delta functions. The delta functions will be eliminated by applying substitution over the $N$ diagonal elements $B_{ii}$ given by the transformation
\begin{equation}
B_{ii} = (x_i - y_i - \sum_{j,j\ne i}B_{ij}y_j)/y_i \label{eq:B_inverse_y}.
\end{equation}
Then the integral can be rewritten into a form without any diagonal elements $B_{ii}$. This requires several layers of algebraic manipulation as shown in detail going forward.

First, the Jacobian of the transformation $J_{ij}=\frac{\partial B_{ii}}{\partial y_j}$ from $(B_{11},B_{22},\dots,B_{NN})\rightarrow (y_1,y_2,\dots,y_N)$ is given by
\begin{align*}
    {\bf J}=-\begin{bmatrix}
    (B_{11}+1)/y_1 & B_{12}/y_1 & \dots & B_{1N}/y_1\\
    B_{21}/y_2 & (B_{22}+1)/y_2 & \dots & B_{2N}/y_2\\
    \vdots & \vdots & \ddots & \vdots \\
    B_{N1}/y_N & B_{N2}/y_N & \dots & (B_{NN}+1)/y_N
    \end{bmatrix}
\end{align*}
where we only need the absolute value of its determinant for the transformation. Since $y_i$ only appears in the $i$-th row we therefore have
\begin{equation}
||{\bf J}|| = \frac{||{\bf I} + {\bf B}_*||}{|y_1 y_2 \dots y_N|} \label{eq:B_to_Y_Jacobian}
\end{equation}
where $||{\bf I}+{\bf B}_*||$ denotes the absolute determinant of the matrices with the diagonal elements of ${\bf B}$ evaluated at their inverse transformations using Eqn.\ref{eq:B_inverse_y}. We now apply the inverse transformation to the argument of the exponent and rearrange the terms to a quadratic form.

Using the block diagonal form for ${\bf C}$ gives
\begin{align*}
\vec{v}^T {\bf C} \vec{v} = \sum_{i} \vec{v}^{(i)T}{\bf F}^{(i)} \vec{v}^{(i)}
\end{align*}
where this form contains the diagonal variables $B_{ii}$ for which we substitute with Eqn.\ref{eq:B_inverse_y}. Next separate each block partitioning as
\begin{equation*}
    \vec{v}^{(m)} = \vec{a}^{(m)} + \vec{b}^{(m)}
\end{equation*}
where $a^{(m)}_i=B_{im}(1-\delta_{im})$  and $b^{(m)}_i=B_{mm}\delta_{im}$. This grouping will facilitate the substitution and we will seek a new expression given by
\begin{equation*}
    \vec{v}^T {\bf C} \vec{v} = \vec{a}^T {\bf H} \vec{a} + \vec{\gamma}^T\vec{a} + \kappa
\end{equation*}
where all expressions on the right-hand side does not contain any diagonal elements $B_{ii}$.

It follows then that
\begin{align*}
\vec{v}^T{\bf C}\vec{v} &= \sum_i (\vec{a}^{(i)}+\vec{b}^{(i)})^T{\bf F}^{(i)}(\vec{a}^{(i)}+\vec{b}^{(i)}) \\
&= \sum_i [ \vec{a}^{(i)T}{\bf F}^{(i)}\vec{a}^{(i)}+\vec{b}^{(i)T}{\bf F}^{(i)}\vec{b}^{(i)}+2\vec{b}^{(i)T}{\bf F}^{(i)}\vec{a}^{(i)}].
\end{align*}
Therefore, the substitution using Eqn.\ref{eq:B_inverse_y} will require manipulating the last two terms since the first term is already in the correct format. The algebraic steps involved can be found in Appendix \ref{app:yalgebra}. The final result gives
\begin{equation}
    H^{(li)}_{jk} = F^{(i)}_{jk}\left[\delta_{li}-2\frac{y_l}{y_i}\delta_{ij}\right] +F^{(j)}_{jj}\delta_{jk}\frac{y_l y_i}{y_j^2}, \label{eq:H}
\end{equation}
\begin{equation}
    \gamma_q^{(i)}=2\left[\beta_i F^{(i)}_{iq} - \frac{\beta_q y_i}{y_q}F^{(q)}_{qq}\right] \label{eq:gamma},
\end{equation}
and
\begin{equation}
    \kappa =\sum_{ik } \frac{A^T_{ik}A_{ki}}{\sigma^2_{ki}}\frac{(x_i-y_i)^2}{y_i^2}.\label{eq:kappa}
\end{equation}
The quadratic coefficient matrix can then be made symmetric as ${\bf H_S} = \frac{1}{2}({\bf H^T} + {\bf H})$.

The final step is to reduce the size of ${\bf H_S}$ from $N^2\times N^2$ to $(N^2-N)\times (N^2-N)$ by removing all rows and columns corresponding to the diagonal locations $B_{ii}$. Similarly, this must be done for all vectors to shrink them from $N^2$ to $(N^2-N)$ elements. Calling the truncated parameters as ${\bf H_S}\rightarrow{\bf \Gamma}$, $\vec{\gamma}\rightarrow\vec{\psi}$, $\vec{a}\rightarrow\vec{\omega}$ gives
\begin{equation*}
    \vec{v}^T {\bf C} \vec{v} = \vec{\omega}^T {\bf \Gamma} \vec{\omega} + \vec{\psi}^T\vec{\omega} + \kappa. 
\end{equation*}
Therefore, after integrating over the $N$ delta functions, the integral for the $\vec{y}$ PDF is given by
\begin{align}
    h_{{\bf A},\vec{x}}(\vec{y},\sigma)=&\frac{\left|\left|{\bf A}\right|\right|^N (2\pi)^{\frac{-N^2}{2}}}{\prod_{ij}| \sigma_{ij}|}\int\frac{||{\bf I}+{\bf \Omega_*}||}{|y_1||y_2|\dots|y_N|} \times\nonumber\\ &\exp\left(\frac{-\vec{\omega}^T {\bf \Gamma} \vec{\omega} - \vec{\psi}^T\vec{\omega} - \kappa}{2}\right)d\vec{\omega}\label{eq:int_y}
\end{align}
where $d\vec{\omega}$ only contains the off-diagonal elements $B_{ij}$ for $i\neq j$, the matrix ${\bf B}\rightarrow{\bf \Omega}$ to be consistent with the removal of diagonal elements of ${\bf B}$, and Eqn.\ref{eq:B_to_Y_Jacobian} gives the Jacobian for which the diagonal elements of ${\bf \Omega_*}$ are evaluated based on Eqn.\ref{eq:B_inverse_y}.

The argument of the exponential can be put into a Gaussian format by completing the square using
\begin{equation}
    \vec{\mu} = -\frac{1}{2}{\bf \Gamma^{-1}}\vec{\psi} \label{eq:mu}
\end{equation}
and
\begin{equation}
    \zeta = \kappa - \vec{\mu}^{T}{\bf \Gamma}\vec{\mu}.\label{eq:zeta}
\end{equation}
It then becomes
\begin{equation*}
    \vec{\omega}^T {\bf \Gamma} \vec{\omega} + \vec{\psi}^T\vec{\omega} + \kappa = (\vec{\omega}-\vec{\mu})^T{\bf \Gamma}(\vec{\omega}-\vec{\mu})+\zeta.
\end{equation*}
Which gives Eqn.\ref{eq:int_y} as
\begin{align}
    h_{{\bf A},\vec{x}}(\vec{y},\sigma)=&\frac{\left|\left|{\bf A}\right|\right|^N (2\pi)^{\frac{-N^2}{2}}}{\prod_{ij}| \sigma_{ij}|}e^{\frac{-\zeta}{2}}\int\frac{||{\bf I}+{\bf \Omega_*}||}{|y_1||y_2|\dots|y_N|} \times\nonumber\\ &\exp\left(-\frac{(\vec{\omega}-\vec{\mu})^T{\bf \Gamma}(\vec{\omega}-\vec{\mu})}{2}\right)d\vec{\omega}.\label{eq:int_gy}
\end{align}
The integral is just the expectation value of $||{\bf I}+{\bf \Omega_*}||$ for a normally distributed random variable $\vec{\omega} \sim \mathcal{N}(\vec{\mu}, {\bf \Gamma}^{-1})$. Conisdering that, we can write the final form of the $\vec{y}$ PDF as

\begin{align}
    h_{{\bf A},\vec{x}}(\vec{y},\sigma)=&\frac{\left|\left|{\bf A}\right|\right|^N (2\pi)^{\frac{-N}{2}}}{\prod_{ij}| \sigma_{ij}|}e^{\frac{-\zeta}{2}}\sqrt{|{\bf \Gamma}^{-1}}| \times\nonumber \\ &
    \frac{\left\langle||{\bf I}+{\bf \Omega_*}||\right\rangle_{\vec{\omega}}}{|y_1||y_2|\dots|y_N|}
\label{eq: analytical_formula_w_expect}
\end{align} 

\subsection{Scaling of the Marginal Distribution $h(y_i)$}
To determine the asymptotic behavior of the marginal distribution $h(y_i)$ for a single species $i$ in the limit $y_i \to \infty$, we start from the scaling property of the joint equilibrium distribution Eq.\ref{eq: analytical_formula_w_expect}. The joint probability density function $h(\vec{y})$ is asymptotically homogeneous of degree $-(N+1)$
\begin{equation}
    h(\lambda \vec{y}) \sim \lambda^{-(N+1)} h(\vec{y}), \quad \text{as } \lambda \to \infty.
    \nonumber
    \label{eq:joint_scaling}
\end{equation}
The marginal distribution is obtained by integrating the joint density over the abundances of all other $N-1$ species coordinates, denoted by the vector $\vec{y}_{- i}$:
\begin{equation}
    h(y_i) = \int_{\mathbb{R}^{N-1}_+} h(\vec{y}) \, d\vec{y}_{- i}.
    \nonumber
\end{equation}
To isolate the dependence on the scale $y_i$, we perform a change of variables to dimensionless relative abundances. Let $\vec{u} = \vec{y}_{- i} / y_i$, such that $\vec{y}_{- i} = y_i \vec{u}$. The differential volume element transforms according to:
\begin{equation}
    d\vec{y}_{- i} = \prod_{j \neq i} d(y_i u_j) = y_i^{N-1} \, d\vec{u}.
    \nonumber
\end{equation}
Substituting the change of variables into the integral and applying the asymptotic homogeneity property from Eq.~\eqref{eq:joint_scaling}
\begin{align}
    h(y_i) &= \int h(y_i, y_i \vec{u}) \, y_i^{N-1} d\vec{u} \nonumber \\
    &\sim \int \left[ y_i^{-(N+1)} h(1, \vec{u}) \right] \, y_i^{N-1} d\vec{u} \nonumber \\
    &\sim y_i^{-(N+1) + (N-1)} \int h(1, \vec{u}) \, d\vec{u}.
\end{align}
The remaining integral over $\vec{u}$ depends only on the relative structure of the ecosystem and converges to a direction independent constant. Thus, the marginal distribution scales as
\begin{equation}
    h(y_i) \sim y_i^{-2}.
\end{equation}

\begin{figure*}
    \centering
    \includegraphics[width=\linewidth]{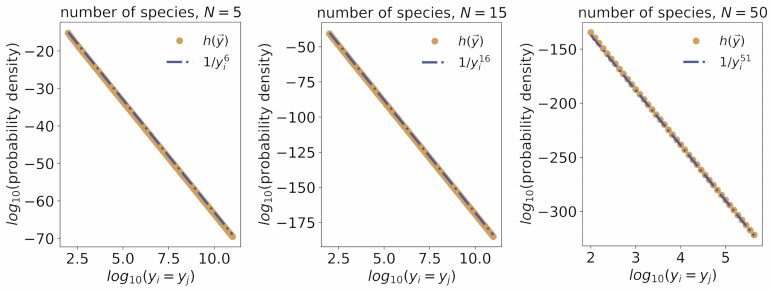}
    \caption{\textbf{Equilibrium abundance probability distribution $h(\vec{y})$ scaling along the main diagonal.} Here we show the universal scaling of the distribution tails with power law $1/y^{N+1}$. From left to right for increasing ecosystem sizes we observe that the scaling is preserved.}
    \label{fig:placeholder}
\end{figure*}

\subsection{Saddle point approximation of the equilibrium distribution}

For communities of arbitrary size $N$ the determinant function is not a linear function of $\vec{w}$ and therefore can be difficult to evaluate. In this case, we can applying the saddle point method in order to get an approximation for this expectation value when the noise is small.


The saddle point method can be applied when the exponential is narrow and centered at the saddle point $\vec{\omega}=-\vec{\mu}$. The $\vec{y}$ integral is of the form
\begin{equation}
    I = \int c(\vec{\omega}) \exp[-f(\vec{\omega})] d\vec{\omega}\label{eq:saddle_form}
\end{equation}
can then be approximated as
\begin{equation*}
    I \approx \int [c(\vec{\mu}) + O(\vec{w}-\vec{\mu})]\exp[-f(\vec{\omega})]d\vec{\omega}
\end{equation*}
where $O(\vec{w}-\vec{\mu})$ are the linear terms in the Taylor series for the determinant of $|{\bf I} + {\bf \Omega}_*|$ with diagonal values given by Eqn.\ref{eq:B_inverse_y} and then taken at $\vec{\omega}=-\vec{\mu}$. These and higher order terms can be determined by following approximations for the determinant given by \cite{ipsen2011determinant}. Decomposing ${\bf I} + {\bf \Omega}_* = ({\bf I}+{\bf D}_* + {\bf M})$ into the diagonal matrix ${\bf I} + {\bf D}_*$ and off-diagonal matrix ${\bf M}$ gives the determinant approximation
\begin{equation*}
    ||{\bf I} + {\bf \Omega}_*|| \approx ||{\bf I} + {\bf D}_*|| \exp\left(Tr\left[({\bf I}+{\bf D}_*)^{-1}{\bf M}\right]\right)
\end{equation*}
when the spectral radius of $([{\bf I}+{\bf D}_*]^{-1}{\bf M})$ is less than 1, and $({\bf I}+{\bf D}_*)$ is non-singular. This approximation can be directly evaluated because the trace of a diagonal matrix ${\bf I}+{\bf D_*}$ multiplying an off-diagonal matrix ${\bf M}$ is zero. The determinant of a diagonal matrix is just the product of its diagonal values
\begin{align}
    ||{\bf I}+{\bf D_*}||&=\prod_{i}\left|1+\frac{x_i-y_i-\sum_j \Omega_{ij}y_j}{y_i}\right|\nonumber\\
    &=\prod_{i}\left|\frac{x_i-\sum_j \Omega_{ij}y_j}{y_i}\right|\label{eq:abs_det}.
\end{align}
The spectral radius condition $\rho([{\bf I}+{\bf D}_*]^{-1}{\bf M})<1$ and non-singular condition are both more easily (but not always) satisfied when the values of the diagonal matrix ${\bf I} + {\bf D_*}$ are large relative to the values of the off-diagonal ${\bf M}$. This is the case whenever the random distribution is sampling nearby values for $\vec{y}=\vec{x}+\vec{\epsilon}$. As $\vec{\epsilon}\rightarrow0$ the random distribution for ${\bf \Omega}$ must sample values near zero. This causes the off-diagonal terms in ${\bf \Omega}$ to be small relative to the diagonal terms given in Eqn.\ref{eq:abs_det}. In this case, the matrix approaches a fully diagonal matrix and therefore both the spectral radius and non-singular condition will be satisfied.

Finally, Eqn.\ref{eq:saddle_form} calculates the expectation value of Eqn.\ref{eq:abs_det} over the distribution of ${\bf \Omega}$. The saddle point method assumes that the exponential is a narrow peak located at $\vec{\omega}=\vec{\mu}$ and applies the approximation $\langle c(\vec{\omega})\rangle\approx c(\langle\vec{\omega}\rangle)$. This gives the final result
\begin{align}
h_{{\bf A},\vec{x}}(\vec{y},\sigma) &\approx \frac{||{\bf A}||^N (2\pi )^{\frac{-N}{2}}}{\prod_{ij}|\sigma_{ij}|} \frac{||{\bf I}+{\bf D}_*||_{\vec{\omega}=\vec{\mu}} e^{\frac{-\zeta}{2}}}{|y_1 y_2\dots y_N|\sqrt{\left|{\bf \Gamma_S}\right|}}\label{eq:y_saddle}
\end{align}
where the expectation value of the absolute determinant is simply Eqn.\ref{eq:abs_det} evaluated at $\vec{\omega}=\vec{\mu}$.

\subsection{Gaussian approximation of the equilibrium distribution}
\label{sec: gaussian_approx.}
We can linearize Eqn. \ref{eq: stochastic_sim_equation} for small noise in {\bf A} to generate a Gaussian distribution centered around $\vec{y}=\vec{x}$. The linear combination of random variables is used to determine the Gaussian moments for $\vec{y}$.

First, taking the linearization around $\vec{y}=\vec{x}$ and assuming that the random elements of B are small relative to the identity matrix ($|{\bf B}|\ll|{\bf I}|$) gives
\begin{equation*}
    \vec{y}\approx [{\bf I}-{\bf B}]\vec{x}
\end{equation*}
where each element of $\vec{y}$ is just a linear combination of the elements of ${\bf B}$ given by
\begin{equation*}
    y_i = x_i - \sum_k B_{ik}x_k.
\end{equation*}
The Gaussian distribution generated from this dependence is easily derived. Its mean is given by
\begin{align*}
    \langle y_i \rangle&=\left\langle x_i - \sum_k B_{ik}x_k\right\rangle\\
    &=x_i
\end{align*}
because $\langle B_{ij} \rangle=0$. The covariance is then given by
\begin{align*}
    Cov(y_i,y_j)&=\langle(y_i-x_i)(y_j-x_j)\rangle\\
    &=\sum_{km}\left\langle B_{ik}B_{jm}\right\rangle x_k x_m
\end{align*}
where $\langle B_{ik}B_{jm}\rangle$ is the covariance between $B_{ik}$ and $B_{jm}$ because they have zero means. This covariance is given by
\begin{equation*}
    Cov(B_{ik},B_{jm})=\sum_p R_{ip}R_{jp}\sigma^2_{pk}\delta_{km}
\end{equation*}
where ${\bf R}={\bf A^{-1}}$. The covariance of $\vec{y}$ is simplified to
\begin{equation}
\label{eq: cov_gaussian}
\Sigma_{\vec{y}}\equiv Cov(y_i,y_j)=\sum_{kp} x_k^2\sigma^2_{pk} R_{ip} R_{jp}
\end{equation}
The average $\vec{y}$ for this approximation is always $\vec{x}$. Going to the second-order approximation of the matrix inverse
\begin{equation*}
    \vec{y}\approx [{\bf I}-{\bf B}+{\bf B^2}]\vec{x}
\end{equation*}
gives a way to track the average $\vec{y}$ as a function of the covariance of ${\bf B}$. This average is given by
\begin{align}
    \langle y_i\rangle&=\langle x_i\rangle+\left\langle\sum_{kj}B_{ik}B_{kj}x_j\right\rangle\nonumber\\
    &=x_i+\sum_{kj}Cov(B_{ik},B_{kj})x_j\nonumber\\
    &=x_i+\sum_{kj} R_{ij}R_{kj}\sigma_{jk}^2x_k.
    \label{eq:gaussian_approximation_mean_second_order}
\end{align}
Defining
\begin{equation*}
    {\bf F} = {\bf R}{\bf R^T}\circ{\bf S}\circ{\bf S}
\end{equation*}
where $\circ$ is the Hadamard product gives the vector form
\begin{equation}
    \langle \vec{y}\rangle=\vec{x}+{\bf F}\vec{x}.
\end{equation}
As expected, when there is no noise ($\sigma\to0$), the distribution is sharply peaked around the original equilibrium position $x$. As we increase the noise, this peak will start departing from its original position, by an amount equal to the second term in eqn(\ref{eq:gaussian_approximation_mean_second_order}). As the noise continues to increase, the expected abundance of one of the species will hit zero. To get a sense of what noise level will cause this, we set all interaction fluctuations equal, and get
\[\sigma=\sqrt{\frac{-x_i}{\sum_{kj} R_{ij}R_{kj}x_k}},\]
where $i$ is the index that gives the smallest real value on the right hand side. If the inside of the square root is negative for a given $i$, this indicates that the expected abundance of $i$ increases with noise, and will therefore not hit zero.

When the spectral radius is small higher order terms of normal distribution correction is negligible.

\subsection{Derivation of TC=CT=I}
\label{app:TC}
Given two block diagonal matrices ${\bf T}$ and ${\bf C}$ the verification that ${\bf TC}={\bf CT}={\bf I}$ can be done separately for each block. If ${\bf AR}={\bf RA}={\bf I}$, then for the $m$-th block observe that
\begin{align*}
    T^{(m,m)}_{ik} C^{(m,m)}_{kj} &= \sum_k\left(\sum_p R_{ip}R^T_{pk}\sigma_{pm}^2\right)\left(\sum_q \frac{A^T_{kq}A_{qj}}{\sigma^2_{qm}}\right)\\
    &= \sum_{pq}\frac{\sigma^2_{pm}}{\sigma^2_{qm}}R_{ip}A_{qj}\sum_k R^T_{pk}A^T_{kq} \\
    &= \sum_{pq}\frac{\sigma^2_{pm}}{\sigma^2_{qm}}R_{ip}A_{qj} \delta_{pq} \\
    &= \sum_{q}\frac{A_{qj}}{\sigma^2_{qm}}\sum_p R_{ip}\sigma_{pm}^2 \delta_{pq} \\
    &= \sum_{q} R_{iq}A_{qj}\frac{\sigma^2_{qm}}{\sigma^2_{qm}} = \delta_{ij}
\end{align*}
which completes the proof. The same result can be shown for $C_{ik}^{(m,m)}T_{kj}^{(m,m)}$.

\subsection{Algebra for the $\vec{y}$ integral}
\label{app:yalgebra}

Starting with the argument of the exponential for the ${\bf B}$ PDF
\begin{align}
\vec{v}^T{\bf C}\vec{v} = &\sum_i (\vec{a}^{(i)}+\vec{b}^{(i)})^T{\bf F}^{(i)}(\vec{a}^{(i)}+\vec{b}^{(i)}) \nonumber \\
=  &\sum_i \vec{a}^{(i)T}{\bf F}^{(i)}\vec{a}^{(i)} \label{eq:app_aFa}\\
+ &\sum_i\vec{b}^{(i)T}{\bf F}^{(i)}\vec{b}^{(i)} \label{eq:app_bFb}\\
+2 &\sum_i\vec{b}^{(i)T}{\bf F}^{(i)}\vec{a}^{(i)} \label{eq:app_bFa}
\end{align}
where Eqn.\ref{eq:app_aFa} is already in the desired format. The substitutions using Eqn.\ref{eq:B_inverse_y} for $N$ diagonal elements of $B_{ii}$ are then given by $b^{(i)}_i=\beta_i - \frac{1}{y_i}\sum_j a^{(j)}_i y_j$ with $\beta_i = \frac{x_i - y_i}{y_i}$. Then Eqn.\ref{eq:app_bFb} becomes
\begin{align}
    \sum_i\vec{b}^{(i)T}{\bf F}^{(i)}\vec{b}^{(i)} =& \sum_i F^{(i)}_{ii}(\beta_i - \frac{1}{y_i}\sum_j a^{(j)}_i y_j)^2 \nonumber \\
    =& \sum_i F^{(i)}_{ii} \beta_i^2 \label{eq:app_betaFbeta}\\
    +& \sum_i \frac{F^{(i)}_{ii}}{y_i^2} (\sum_j a^{(j)}_i y_j)^2 \label{eq:app_Fsuma2} \\
    -2& \sum_i \frac{F^{(i)}_{ii}\beta_i}{y_i} \sum_j a^{(j)}_i y_j. \label{eq:app_betaFsuma}
\end{align}
The first and third term requires no additional manipulation. The second term given by Eqn.\ref{eq:app_Fsuma2} can be rewritten as
\begin{align}
    \sum_i \frac{F^{(i)}_{ii}}{y_i^2} (\sum_j a^{(j)}_i y_j)^2 =\sum_{imq} a_i^{(m)}F^{(i)}_{ii}\frac{y_m y_q}{y_i^2} a_i^{(q)}. \label{eq:app_Famaq}
\end{align}
Going back to Eqn.\ref{eq:app_bFa} and applying the substitution gives
\begin{align}
    2 \sum_i\vec{b}^{(i)T}{\bf F}^{(i)}\vec{a}^{(i)} = 2&\sum_{iq} (\beta_i - \frac{1}{y_i}\sum_j a^{(j)}_i y_j)F^{(i)}_{iq}a^{(i)}_q\nonumber\\
    = 2&\sum_{iq} \beta_iF^{(i)}_{iq}a_q^{(i)} \label{eq:app_betaFa}\\
    -&\sum_{iq}\frac{2}{y_i}\sum_j a^{(j)}_i y_j F_{iq}^{(i)} a_q^{(i)} \label{eq:app_sumaFsuma}.
\end{align}
At this point all terms substitutions have been completed. The final task is to group quadratic, linear, and constant terms with respect to $\vec{a}$. The constant term is given by Eqn.\ref{eq:app_betaFbeta}
\begin{equation}
    \kappa = \sum_i F_{ii}^{(i)}\beta_i^2=\sum_{ik } \frac{A^T_{ik}A_{ki}}{\sigma^2_{ki}}\frac{(x_i-y_i)^2}{y_i^2}.
\end{equation}

The linear terms are given by Eqn.\ref{eq:app_betaFsuma} and Eqn.\ref{eq:app_betaFa}. They should be rearranged to match the form $\vec{\gamma}^T\vec{a}$ to get
\begin{align*}
    2&\sum_{iq} \beta_i F^{(i)}_{iq}a_q^{(i)} -2 \sum_i \frac{F^{(i)}_{ii}\beta_i}{y_i} \sum_j a^{(j)}_i y_j= \\
    2&\sum_{iq} [\beta_i F^{(i)}_{iq} - \frac{\beta_q y_i}{y_q}F^{(q)}_{qq}] a^{(i)}_q.
\end{align*}
Therefore the linear vector is
\begin{equation}
    \gamma_q^{(i)}=2\left[\beta_i F^{(i)}_{iq} - \frac{\beta_q y_i}{y_q}F^{(q)}_{qq}\right].
\end{equation}
Finally, the quadratic terms given by Eqn.\ref{eq:app_aFa}, \ref{eq:app_Famaq}, and \ref{eq:app_sumaFsuma} must be rearranged to match $\vec{a}^T{\bf H}\vec{a}$ to get
\begin{align*}
    \sum_{ijkl} a^{(l)}_j H_{jk}^{(li)} a^{(i)}_k = &\sum_{ijkl} a_j^{(l)} F^{(i)}_{jk}\delta_{li} a_k^{(i)} \\
    +&\sum_{ijkl} a_j^{(l)}F^{(j)}_{jj}\delta_{jk}\frac{y_l y_i}{y_j^2} a_k^{(i)}\\
    -&\sum_{ijkl}a_j^{(l)}\frac{2F^{(j)}_{jk}y_l}{y_i}\delta_{ij}a_k^{(i)}.
\end{align*}
Therefore the quadratic coefficient matrix is given by
\begin{equation}
    H^{(li)}_{jk} = F^{(i)}_{jk}\left[\delta_{li}-2\frac{y_l}{y_i}\delta_{ij}\right] +F^{(j)}_{jj}\delta_{jk}\frac{y_l y_i}{y_j^2}
\end{equation}

\section{Explicit formulas for a two species community}
A two species system where $N=2$ is defined by
\begin{equation*}
    {\bf A} = \begin{bmatrix}
    A_{11} & A_{12} \\
    A_{21} & A_{22}
    \end{bmatrix},
    \sigma = \begin{bmatrix}
    \sigma_{11} & \sigma_{12} \\
    \sigma_{21} & \sigma_{22}
    \end{bmatrix},
    \vec{x}=\begin{bmatrix}
    x_1 \\
    x_2
    \end{bmatrix}
\end{equation*}
which respectively describes the mean interactions, standard deviation in the interaction noise, and initial equilibrium in absence of noise. The intrinsic vector $\vec{r}=-{\bf A}\vec{x}$ can be determined based on the given values. The PDF for $\vec{y}=[{\bf I}+{\bf A^{-1}W}]\vec{x}$ will be evaluated by following the procedure described above.

The first step is to determine the general form for $g({\bf B},\sigma)$ where ${\bf B}={\bf A^{-1}W}$. This is given by Eqn.\ref{eq:B_pdf} using Eqn.\ref{eq:C_definition} and Eqn.\ref{eq:F_definition}. First, the indexing for the $N^2=4$ dimension must be arranged. Following the same partitioning as above gives
\begin{equation*}
    \vec{v}=\begin{bmatrix}
    \vec{v}^{(1)}\\
    \hline
    \vec{v}^{(2)}
    \end{bmatrix}
    =\begin{bmatrix}
    B_{11}\\
    B_{21}\\
    \hline
    B_{12}\\
    B_{22}
    \end{bmatrix}.
\end{equation*}
There are $N=2$ block matrices given by Eqn.\ref{eq:F_definition} as
\begin{align*}
    {\bf F}^{(1)} &= \begin{bmatrix}
    \frac{A^T_{11}A_{11}}{\sigma^2_{11}}+\frac{A^T_{12}A_{21}}{\sigma^2_{21}} && \frac{A^T_{11}A_{12}}{\sigma^2_{11}}+\frac{A^T_{12}A_{22}}{\sigma^2_{21}}\\
    \frac{A^T_{21}A_{11}}{\sigma^2_{11}}+\frac{A^T_{22}A_{21}}{\sigma^2_{21}} && \frac{A^T_{21}A_{12}}{\sigma^2_{11}}+\frac{A^T_{22}A_{22}}{\sigma^2_{21}}
    \end{bmatrix},\\
    {\bf F}^{(2)} &= \begin{bmatrix}
    \frac{A^T_{11}A_{11}}{\sigma^2_{12}}+\frac{A^T_{12}A_{21}}{\sigma^2_{22}} && \frac{A^T_{11}A_{12}}{\sigma^2_{12}}+\frac{A^T_{12}A_{22}}{\sigma^2_{22}}\\
    \frac{A^T_{21}A_{11}}{\sigma^2_{12}}+\frac{A^T_{22}A_{21}}{\sigma^2_{22}} && \frac{A^T_{21}A_{12}}{\sigma^2_{12}}+\frac{A^T_{22}A_{22}}{\sigma^2_{22}}
    \end{bmatrix},
\end{align*}
and Eqn.\ref{eq:C_definition} gives the block matrix form for the precision matrix as
\begin{equation*}
    {\bf C}=\begin{bmatrix}
    {\bf F}^{(1)} & {\bf 0}\\
    {\bf 0} & {\bf F}^{(2)}
    \end{bmatrix}
\end{equation*}
where ${\bf 0}$ is a $N\times N$ matrix of zeroes. Therefore, the PDF for the distribution of ${\bf B}={\bf A^{-1}}{\bf W}$ from Eqn.\ref{eq:B_pdf} is
\begin{align}
    g({\bf B},\sigma) =& \frac{\left|{ A_{11}A_{22}-A_{12}A_{21}}\right|^2}{(2\pi)^{2}| \sigma_{11}\sigma_{12}\sigma_{21}\sigma_{21}|} \exp(-\frac{\vec{v}^T {\bf C} \vec{v}}{2}) \label{eq:B_pdf_N2}
\end{align}
where the argument inside the exponential is
\begin{align*}
    &\vec{v}^T{\bf C}\vec{v}=\begin{bmatrix}
    B_{11} & B_{21}
    \end{bmatrix}{\bf F}^{(1)}\begin{bmatrix}
    B_{11} \\
    B_{21}
    \end{bmatrix} + \begin{bmatrix}
    B_{12} & B_{22}
    \end{bmatrix}{\bf F}^{(2)}\begin{bmatrix}
    B_{12} \\
    B_{22}
    \end{bmatrix}\\
    &=\left(\frac{A^T_{11}A_{11}}{\sigma^2_{11}}+\frac{A^T_{12}A_{21}}{\sigma^2_{21}}\right)B_{11}^2+\left(\frac{A^T_{21}A_{12}}{\sigma^2_{11}}+\frac{A^T_{22}A_{22}}{\sigma^2_{21}}\right)B_{21}^2\\
    &+2\left(\frac{A^T_{11}A_{12}}{\sigma^2_{11}}+\frac{A^T_{12}A_{22}}{\sigma^2_{21}}\right)B_{11}B_{21}\\
    &+\left(\frac{A^T_{11}A_{11}}{\sigma^2_{12}}+\frac{A^T_{12}A_{21}}{\sigma^2_{22}}\right) B_{12}^2+\left(\frac{A^T_{21}A_{12}}{\sigma^2_{12}}+\frac{A^T_{22}A_{22}}{\sigma^2_{22}}\right) B_{22}^2 \\
    &+2\left(\frac{A^T_{11}A_{12}}{\sigma^2_{12}}+\frac{A^T_{12}A_{22}}{\sigma^2_{22}}\right) B_{12}B_{22}.
\end{align*}

Next, the integral for $\vec{y}$ PDF is given by Eqn.\ref{eq:int_y} which requires the Jacobian in Eqn.\ref{eq:B_to_Y_Jacobian} along with the parameters given in Eqns.\ref{eq:H}, \ref{eq:gamma}, and \ref{eq:kappa}. The Jacobian requires evaluating the diagonal elements of ${\bf I} + {\bf B}$ using the substitution Eqn.\ref{eq:B_inverse_y}. Doing this gives
\begin{equation}
    {\bf B}_*=\begin{bmatrix}
    (x_1-y_1-B_{12}y_2)/y_1 & B_{12}\\
    B_{21} & (x_2-y_2-B_{21}y_1)/y_2
    \end{bmatrix}\label{eq:B*N2}
\end{equation}
for which the absolute determinant in Eqn.\ref{eq:B_to_Y_Jacobian} becomes
\begin{align*}
    ||{\bf I}+{\bf B}_*||=\bigg|&\frac{x_1 x_2-B_{21}x_1 y_1-B_{12}x_2 y_2}{y_1 y_2}\bigg|.
\end{align*}
Then ${\bf H}$ in the form
\begin{equation*}
    {\bf H}=\begin{bmatrix}
    H_{11}^{(11)} & H_{12}^{(11)} \vline & H_{11}^{(12)} & H_{12}^{(12)} \\
    H_{21}^{(11)} & H_{22}^{(11)} \vline & H_{21}^{(12)} & H_{22}^{(12)} \\
    \hline
    H_{11}^{(21)} & H_{12}^{(21)} \vline & H_{11}^{(22)} & H_{12}^{(22)} \\
    H_{21}^{(21)} & H_{22}^{(21)} \vline & H_{21}^{(22)} & H_{22}^{(22)} 
    \end{bmatrix}
\end{equation*}
will have its first and last row and column removed due to the ordering of $\vec{v}$. Therefore, the only required elements are given by Eqn.\ref{eq:H} as
\begin{align*}
    H_{22}^{(11)} &= F_{22}^{(1)}+F_{22}^{(2)}\frac{y_1^2}{y_2^2}\\
    H_{21}^{(12)} &= -2F_{21}^{(2)}\frac{y_1}{y_2}\\
    H_{12}^{(21)} &= -2F_{12}^{(1)}\frac{y_2}{y_1}\\
    H_{11}^{(22)} &= F_{11}^{(2)}+F_{11}^{(1)}\frac{y_2^2}{y_1^2}.
\end{align*}
Similarly, the required elements of $\vec{\gamma}$ are given by Eqn.\ref{eq:gamma} as
\begin{align*}
    \gamma_2^{(1)} &= 2\left[\frac{x_1-y_1}{y_1}F_{12}^{(1)}-\frac{(x_2-y_2)y_1}{y_2^2}F_{22}^{(2)}\right]\\
    \gamma_1^{(2)} &= 2\left[\frac{x_2-y_2}{y_2}F_{21}^{(2)}-\frac{(x_1-y_1)y_2}{y_1^2}F_{11}^{(1)}\right],
\end{align*}
and Eqn.\ref{eq:kappa} gives
\begin{align*}
    \kappa = F_{11}^{(1)}\frac{(x_1-y_1)^2}{y_1^2} + F_{22}^{(2)}\frac{(x_2-y_2)^2}{y_2^2}.
\end{align*}
The removal of the indicies corresponding to the diagonals $B_{11}$ and $B_{22}$ gives the reduction $\vec{v}\rightarrow\vec{\omega}$
\begin{equation*}
    \vec{\omega}=\begin{bmatrix}
    B_{21}\\
    B_{12}
    \end{bmatrix},
\end{equation*}
$\vec{\gamma}\rightarrow\vec{\psi}$ as
\begin{equation*}
    \vec{\psi}=2\begin{bmatrix}
    \frac{x_1-y_1}{y_1}F_{12}^{(1)}-\frac{(x_2-y_2)y_1}{y_2^2}F_{22}^{(2)} \\
    \frac{x_2-y_2}{y_2}F_{21}^{(2)}-\frac{(x_1-y_1)y_2}{y_1^2}F_{11}^{(1)}
    \end{bmatrix},
\end{equation*}
and the reduced symmetric matrix is given by ${\bf H_S}\rightarrow{\bf \Gamma}$
\begin{equation*}
    {\bf \Gamma}=\begin{bmatrix}
    F_{22}^{(1)}+F_{22}^{(2)}\frac{y_1^2}{y_2^2} & -F_{21}^{(2)}\frac{y_1}{y_2}-F_{12}^{(1)}\frac{y_2}{y_1} \\
    -F_{21}^{(2)}\frac{y_1}{y_2}-F_{12}^{(1)}\frac{y_2}{y_1}  & F_{11}^{(2)}+F_{11}^{(1)}\frac{y_2^2}{y_1^2}
    \end{bmatrix}.
\end{equation*}
These parameters together gives the integral for the $\vec{y}$ PDF in Eqn.\ref{eq:int_y} as
\begin{align*}
    h_{{\bf A},\vec{x}}(\vec{y},\sigma)=\frac{\left|{ A_{11}A_{22}-A_{12}A_{21}}\right|^2}{(2\pi)^{2}| \sigma_{11}\sigma_{12}\sigma_{21}\sigma_{22}||y_1 y_2|^2} \times \nonumber\\
    \int\bigg|x_1 x_2-\omega_1 x_1 y_1-\omega_2 x_2 y_2\bigg| \times\nonumber\\ \exp\left(\frac{-\vec{\omega}^T {\bf \Gamma} \vec{\omega} - \vec{\psi}^T\vec{\omega} - \kappa}{2}\right)d\omega_1 d\omega_2.
\end{align*}
The next step requires completing the square for the argument inside the exponential using Eqn.\ref{eq:mu}
\begin{equation*}
    \vec{\mu}=\frac{-1}{2 |{\bf \Gamma}|}\begin{bmatrix}
    F_{11}^{(2)}+F_{11}^{(1)}\frac{y_2^2}{y_1^2}& F_{21}^{(2)}\frac{y_1}{y_2}+F_{12}^{(1)}\frac{y_2}{y_1} \\
    F_{21}^{(2)}\frac{y_1}{y_2}+F_{12}^{(1)}\frac{y_2}{y_1}  & F_{22}^{(1)}+F_{22}^{(2)}\frac{y_1^2}{y_2^2}
    \end{bmatrix}\vec{\psi}
\end{equation*}
and Eqn.\ref{eq:zeta}
\begin{equation*}
    \zeta = F_{11}^{(1)}\frac{(x_1-y_1)^2}{y_1^2} + F_{22}^{(2)}\frac{(x_2-y_2)^2}{y_2^2} - \vec{\mu}^{T}{\bf \Gamma}\vec{\mu}.
\end{equation*}
The entire exponential is then separated based on their dependence upon $\omega_1$ and $\omega_2$ as
\begin{equation*}
    e^{\frac{-\vec{\omega}^T {\bf \Gamma} \vec{\omega} - \vec{\psi}^T\vec{\omega} - \kappa}{2}}=e^{\frac{-\zeta}{2}}e^{\frac{-(\vec{\omega}-\vec{\mu})^T{\bf \Gamma}(\vec{\omega}-\vec{\mu})}{2}}.
\end{equation*}
The first exponential can be distributed out of the integral to get the final form
\begin{align}
    h_{{\bf A},\vec{x}}(\vec{y},\sigma)=\frac{\left|{ A_{11}A_{22}-A_{12}A_{21}}\right|^2e^{\frac{-\zeta}{2}}}{(2\pi)^{2}| \sigma_{11}\sigma_{12}\sigma_{21}\sigma_{22}||y_1 y_2|^2} \times \nonumber\\
    \int\bigg|x_1 x_2-\omega_1 x_1 y_1-\omega_2 x_2 y_2\bigg| \times\nonumber\\ \exp\left[\frac{-(\vec{\omega}-\vec{\mu})^T{\bf \Gamma}(\vec{\omega}-\vec{\mu})}{2}\right]d\omega_1 d\omega_2 \label{eq:int_y_N2}.
\end{align}
Although the determinant for the $N=2$ case is a linear function of $\vec{w}$ the absolute value causes this integral to be non-trivial. However, because it is an expectation value of Gaussian random variables it can be efficiently evaluated in the form

\begin{align} \label{expectation_value_PDF_N2}
    h_{{\bf A},\vec{x}}(\vec{y},\sigma)=\sqrt{\left|{\bf \Gamma_S}^{-1}\right|}\frac{\left|{ A_{11}A_{22}-A_{12}A_{21}}\right|^2e^{\frac{-\zeta}{2}}}{2\pi| \sigma_{11}\sigma_{12}\sigma_{21}\sigma_{22}||y_1 y_2|^2} \times \nonumber\\
    \left\langle\bigg|x_1 x_2-\omega_1 x_1 y_1-\omega_2 x_2 y_2\bigg|\right\rangle_{\vec{\omega}}
\end{align}

where the expectation value $\langle\cdot\rangle_{\vec{\omega}}$ is evaluated over numerical samples of $\vec{\omega}$. This numerical evaluation can be compared to the result of the saddle point method to determine the regimes for which the saddle point method is a good approximation.

More specifically, Eqn. \ref{eq:int_gy} transfers the problem of sampling for $\vec{y}$ to sampling for $\vec{\omega}$. This procedure is advantageous when the desired values for $\vec{y}$ have very low probabilities, like when examining the tails of distributions. Additionally, the sampling space for $\vec{y}$ requires generating $N^2$ random variables for ${\bf A}$. In comparison, the sampling space for $\vec{\omega}$ requires $N^2-N$ random variables.

Moving forward, the saddle point method requires approximating the absolute determinant inside the integral as an expansion for deviations $\Delta\vec{w}$ around $\vec{w}=\vec{\mu}$. Then the determinant of the diagonal evaluated at $\vec{w}=\vec{\mu}$ will be used to approximate the expectation value.

Then, following Eqn.\ref{eq:saddle_form} requires letting $f(\vec{\omega})=(\vec{\omega}-\vec{\mu})^T{\bf \Gamma}(\vec{\omega}-\vec{\mu})$ which gives the saddle point as $\vec{\omega}^*=\vec{\mu}$. The absolute determinant $c(\vec{\omega})=||{\bf I} + {\bf \Omega}_*||$ is more easily manipulated near the saddle point for $\vec{\omega} = \vec{\mu} + \Delta\vec{\omega}$
\begin{align*}
    c(\Delta\omega_1,\Delta\omega_2)|_{\vec{\omega}=\vec{\mu}}&=||{\bf I}+
    {\bf \Omega}_*||\\
    &=\begin{Vmatrix}
    \frac{x_1-(\mu_2+\Delta\omega_2 )y_2}{y_1} & \mu_2+\Delta\omega_2\\
    \mu_1+\Delta\omega_1 & \frac{x_2-(\mu_1+\Delta\omega_1) y_1}{y_2}
    \end{Vmatrix}.
\end{align*}
Following the approximation given in \cite{ipsen2011determinant} with the decomposition ${\bf I}+{\bf \Omega}_* = {\bf D} + {\bf M}$ for the diagonal matrix
\begin{equation*}
    {\bf D} = \begin{bmatrix}
    \frac{x_1-(\mu_2+\Delta\omega_2 )y_2}{y_1} & 0\\
    0 & \frac{x_2-(\mu_1+\Delta\omega_1) y_1}{y_2}
    \end{bmatrix}
\end{equation*}
and off-diagonal matrix
\begin{equation*}
    {\bf M} = \begin{bmatrix}
    0 & \mu_2+\Delta\omega_2\\
    \mu_1+\Delta\omega_1 & 0
    \end{bmatrix}
\end{equation*}
gives the first order approximation $O(\Delta\vec{\omega})$ for the determinant as
\begin{equation*}
    c(\Delta\omega_1,\Delta\omega_2)|_{\vec{w}=\vec{\mu}}\approx \left||{\bf D}|\exp\left(\Tr{\bf D^{-1}M}\right)\right|
\end{equation*}
which is valid when the spectral radius $\rho({\bf D^{-1}M})<1$. This can be evaluated to get
\begin{equation}
    c(\Delta\vec{\omega})|_{\vec{w}=\vec{\mu}}\approx \left|\frac{x_1-(\mu_2+\Delta\omega_2 )y_2}{y_1}\frac{x_2-(\mu_1+\Delta\omega_1) y_1}{y_2}\right|\label{eq:c_det_approx}
\end{equation}
because the trace inside the exponential is zero. At this point it is important to go back to the full determinant
\begin{align}
    ||{\bf I} + {\bf \Omega}_*|| = &\bigg|\frac{x_1-(\mu_2+\Delta\omega_2 )y_2}{y_1}\frac{x_2-(\mu_1+\Delta\omega_1) y_1}{y_2}\nonumber\\
    &-(\mu_1+\Delta\omega_1)(\mu_2+\Delta\omega_2)\bigg|\label{eq:c_det_full}
\end{align}
to determine the deviation between Eqn.\ref{eq:c_det_approx} and Eqn.\ref{eq:c_det_full}. This directly shows that $||{\bf I}+{\bf \Omega}_*||\leq|(\Delta\omega_1+\mu_1)(\Delta\omega_2+\mu_2)|+|c(\Delta\vec{\omega})|_{\vec{w}=\vec{\mu}}$. Therefore, the error between the approximation and the true absolute determinant is bounded by $|(\Delta\omega_1+\mu_1)(\Delta\omega_2+\mu_2)|$. Hence, this approximation will work well for samples near $\vec{w}=\vec{\mu}$, especially when $\vec{\mu}\rightarrow0$.

At this point there is a fork in the path towards an analytical expression for the $\vec{y}$ PDF. Applying the generic form of the determinant given in Eqn.\ref{eq:abs_det} gives the result Eqn.\ref{eq:c_det_approx} while applying the more accurate expression for the determinant for $N=2$ gives Eqn.\ref{eq:c_det_full}. We can try both paths and later evaluate which is better.

First, setting $\Delta\vec{\omega}=0$ in Eqn.\ref{eq:c_det_approx} for the determinant in the integral in Eqn.\ref{eq:int_y} and then integrating gives
\begin{align}
    h_{{\bf A},\vec{x}}(\vec{y},\sigma)\approx\frac{\left|{ A_{11}A_{22}-A_{12}A_{21}}\right|^2e^{\frac{-\zeta}{2}}}{2\pi| \sigma_{11}\sigma_{12}\sigma_{21}\sigma_{21}||y_1 y_2|^2} \sqrt{\left|{\bf \Gamma_S}^{-1}\right|} \times \nonumber\\
    \left|x_1-\mu_2y_2||x_2-\mu_1 y_1\right|.\label{eq:y_pdf_saddle_N2}
\end{align}

Second, setting $\Delta\vec{\omega}=0$ in Eqn.\ref{eq:c_det_full} for the determinant in the integral in Eqn.\ref{eq:int_y} and then integrating gives

\begin{align} \label{saddle_point_PDF_N2}
    h_{{\bf A},\vec{x}}(\vec{y},\sigma)\approx\frac{\left|{ A_{11}A_{22}-A_{12}A_{21}}\right|^2e^{\frac{-\zeta}{2}}}{2\pi| \sigma_{11}\sigma_{12}\sigma_{21}\sigma_{21}||y_1 y_2|^2} \sqrt{\left|{\bf \Gamma_S}^{-1}\right|}\times \nonumber\\
    \left|(x_1-\mu_2 y_2)(x_2-\mu_1 y_1)-\mu_1^2\mu_2^2\right|.
\end{align}

Therefore, the location for which $h\rightarrow0$ is different for the two approximations. The less accurate determinant (but more general and easier to write down) incorrectly identifies that $h\rightarrow0$ whenever $\mu_1=x_2/y_1$ or $\mu_2=x_1/y_2$. The more accurate determinant correctly identifies that this location depends upon both $\mu_1$ and $\mu_2$ together $\mu_1^2\mu_2^2=(x_1-\mu_2 y_2)(x_2-\mu_1 y_1)$. However, both cases asymptotically converge  as $\vec{\mu}\rightarrow0$.

\section{Modified Lotka Volterra Model with Holling Type Response Lead to Power Law}
If we modify Generalized Lotka Volterra Model to change all of the interactions to be saturating functions we can write
\begin{equation}
\frac{dn_i}{dt} = n_i\bigg[r_i + \sum_{j=1}^N A_{ij} \phi_j(n_j)\bigg]
\label{eq: generalized_LV}
\end{equation}
governing the abundance $n_i(t)$ of the $i^{\mbox{th}}$ species at time $t$, with intrinsic growth rate  $r_i$. Here $A_{ij}$ is the interaction matrix, and $\phi$ the functional response. Setting the square bracket to zero yields the coexistent equilibrium condition, 
\begin{align}
n_j=\phi_j^{-1}(x_j)\quad \mbox{where} \quad
\vec{x}=-{\bf A}^{-1}\vec{r}
\label{equilibrium}
\end{align}
The quantity $\vec{x}$ uniquely determines the abundances. For example, for a Holling-type functional response with exponent $m_i$, we have $n_i = \beta_i[x_i/(\alpha_i - x_i)]^{1/m_i}$, whereas for the Lotka-Volterra model, $\vec{n}=\vec{x}$. The system is said to be ``feasible'' if $n_i>0$, for all $i$ which for biologically realistic functional responses, is equivalent to $x_i>0$. Even though this model has limitations in its biology as all of the interactions including self limitation has a saturating point, it serves as a good example for a different mechanism that can lead to heavy tailed abundance distributions.

For such a Holling type functional response, abundance distribution for normal Lotka Volterra Model, $h(\vec{y})$, helps us determine the probability distribution of the abundances through
$
p(\vec{n}) = h(\vec{y})\left|\det{({\bf J})}\right|
$
where ${\bf J}$ is a diagonal jacobian matrix and $J_{ii}=\frac{m_in_i^{m-1}\alpha_i\beta_i}{(\beta_i^m + n_i^m)^2}$. Because $n_i\gg\beta_i$, $p(\vec{n})$ scales as a power law,
\begin{equation}
p(\vec{n})\sim m_i\,\alpha_i\,\beta_i^{m_i}\,h(y_1, y_2, ...,\alpha_i,..., y_N)\,n_i^{-(m_i+1)}.
\label{eq:holling_powerlaw_general}
\end{equation}
The tail behavior depends on the direction followed. Depending on that, power of the tails are weaker in the direction of simultaneously increasing $n_i$ due to multiplication of powers $-(m_i+1)$. Heavy tail form of $p(\vec{n})$ is a purely
geometric consequence of mapping a bounded variable
$y\in(0,1)$ onto an unbounded abundance $n\in(0,\infty)$. Even if the equilibrium
distribution of the transformed variable $y_i$ were narrow or Gaussian,
the observable abundances $n_i = \phi_i^{-1}(y_i)$ would still exhibit a
heavy-tailed form. Interestingly, the saturating Holling form that typically stabilizes population dynamics, in this context broadens equilibrium statistics, thereby increasing the risk of rare events. 

An interesting result emerges when we look at the marginal distribution of equilibrium abundances for one species, $p(n_i)$. As $n_i$ depends only on $y_i$, similar to the way we have written multivariate distribution, this marginal distribution is given as $p(n_i)=h(y_i)|\frac{dy_i}{dn_i}|$. For the large values on $n_i$, the marginal distribution $h(y_i)$ approaches to the constant value $h(\alpha_i)$ and the distribution scales with the derivative which we find as

\begin{equation}
    \label{eq: holling_marginal_scaling}
    p(n_i) \sim n_i^{-(m_i+1)}
\end{equation}
This shows that we can expect single species equilibrium abundances to follow power law tails with an exponent dependent on their Holling-type functional response.

Eq. \ref{eq:holling_powerlaw_general} shows that the tail exponent is independent of the detailed form of $h(\vec{y})$. Nevertheless, the distribution $h(\vec{y})$ remains essential for
the finite $\vec{n}$ regime (including how quickly the power law is reached), the anisotropy and cross species covariance of equilibrium shifts. Most importantly, $h(\vec{y})$ is the distribution of equilibrium abundances in the Generalized Lotka Volterra case.

\subsection{Abundance distributions look log-normal in the low noise regime}

Here, we will approximate to the distribution of equilibrium abundances $p(\vec{n})$ when the holling type response governs. Specifically, we will be working in the low noise regime such that equilibrium distribution in generalized Lotka-Volterra system can be approximated to a multivariate Gaussian distribution as we did in Sec. \ref{sec: gaussian_approx.}.

Let $\vec{y}=\vec{x}+\delta  \vec{y}$ with $\mathbb E[\delta \vec{y}]=0$ and ${\bf \Sigma}_{y} \equiv Cov(y_i,y_j)=\sum_{kp} x_k^2\sigma^2_{pk} R_{ip} R_{jp}$, where we assume
$\delta \vec{y}$ is Gaussian.
Writing the Holling map and defining

\begin{align*}
    n_i &= \phi_i^{-1}(y_i)=\beta_i\Big(\frac{y_i}{\alpha_i-y_i}\Big)^{1/m_i} \\
    \vec{z} &= \text{ln}(\vec{n}) = g(\vec{y})
\end{align*}
The vector transformation $\vec{g}(\vec{y})$ is component wise, meaning $z_i = g_i(y_i)$. We can find the explicit form of $g_i$ by taking the natural logarithm of $n_i$:
\begin{equation}
z_i = g_i(y_i) = \ln \beta_i + \frac{1}{m_i} \left[ \ln y_i - \ln(\alpha_i - y_i) \right]
\end{equation}
This is a key simplification as $z_i$ depends only on $y_i$, the Jacobian matrix of $\vec{g}$ will be diagonal, and the Hessian matrix $\mathbf{H}_i$ for each component $g_i$ will also be diagonal.

\subsubsection*{Mean of $\vec{z}$}
To find the mean, we must use a second order Taylor expansion to account for mean shift. The general expansion for a single component $z_i = g_i(\vec{y})$ around the mean $\vec{x}$ is:
$$
z_i = g_i(\vec{x} + \delta\vec{y}) \approx g_i(\vec{x}) + \nabla g_i(\vec{x})^T \delta\vec{y} + \frac{1}{2} \delta\vec{y}^T \mathbf{H}_i(\vec{x}) \delta\vec{y}
$$
where $\nabla g_i$ is the gradient of $g_i$ and $\mathbf{H}_i$ is its Hessian matrix.

To find the mean, we take the expectation of this expansion:
$$
\mu_{z,i} = \mathbb{E}[z_i] \approx \mathbb{E}[g_i(\vec{x})] + \mathbb{E}[\nabla g_i(\vec{x})^T \delta\vec{y}] + \mathbb{E}\left[\frac{1}{2} \delta\vec{y}^T \mathbf{H}_i(\vec{x}) \delta\vec{y}\right]
$$
This gives the general result for the mean:
$$
\mu_{z,i} \approx g_i(\vec{x}) + \frac{1}{2} \operatorname{Tr}\left(\mathbf{H}_i(\vec{x}) \mathbf{\Sigma}_y\right)
$$
Because our transformation is component wise, the Hessian matrix $\mathbf{H}_i$ for $g_i$ is a matrix of zeros except for the diagonals.
$$
(\mathbf{H}_i)_{jk} = \frac{\partial^2 g_i}{\partial y_j \partial y_k} = \begin{cases} g_i''(x_i) & \text{if } j=k=i \\ 0 & \text{otherwise} \end{cases}
$$
The trace term therefore simplifies dramatically:
$$
\operatorname{Tr}(\mathbf{H}_i \mathbf{\Sigma}_y) = \sum_{j,k} (\mathbf{H}_i)_{jk} (\mathbf{\Sigma}_y)_{kj} = (\mathbf{H}_i)_{ii} (\mathbf{\Sigma}_y)_{ii} = g_i''(x_i) \Sigma_{y,ii}
$$
Thus, the $i$-th component of the mean vector $\vec{\mu}_z$ is:
$$
\mu_{z,i} \approx g_i(x_i) + \frac{1}{2} g_i''(x_i) \Sigma_{y,ii}
$$
We calculate the first and second derivatives of $g_i(y_i)$ and evaluate them at the mean $x_i$:
\begin{align*}
    g_i'(y_i) &= \frac{1}{m_i} \left[ \frac{1}{y_i} + \frac{1}{\alpha_i - y_i} \right] \\
    g_i''(y_i) &= \frac{1}{m_i} \left[ -\frac{1}{y_i^2} + \frac{1}{(\alpha_i - y_i)^2} \right]
\end{align*}

\subsubsection*{Covariance of $\vec{z}$}
To find the output covariance matrix $\mathbf{\Sigma}_z$, the first-order expansion is sufficient for the dominant term
$$
\vec{z} = \vec{g}(\vec{x} + \delta\vec{y}) \approx \vec{g}(\vec{x}) + \mathbf{J}_g(\vec{x}) \delta\vec{y}
$$
where $\mathbf{J}_g(\vec{x})$ is the Jacobian matrix of the transformation $\vec{g}$, with entries $(J_g)_{ij} = \frac{\partial g_i}{\partial y_j}$.
The fluctuation of the output is $\delta\vec{z} = \vec{z} - \vec{\mu}_z \approx \mathbf{J}_g(\vec{x}) \delta\vec{y}$. The covariance matrix can be written as:
$$
\mathbf{\Sigma}_z = \mathbb{E}[\delta\vec{z} \delta\vec{z}^T] \approx \mathbb{E}\left[ (\mathbf{J}_g \delta\vec{y}) (\mathbf{J}_g \delta\vec{y})^T \right]
$$
$$
\mathbf{\Sigma}_z \approx \mathbb{E}\left[ \mathbf{J}_g \delta\vec{y} \delta\vec{y}^T \mathbf{J}_g^T \right] = \mathbf{J}_g \mathbb{E}\left[ \delta\vec{y} \delta\vec{y}^T \right] \mathbf{J}_g^T
$$
This gives the general covariance result:
$$
\mathbf{\Sigma}_z \approx \mathbf{J}_g(\vec{x}) \mathbf{\Sigma}_y \mathbf{J}_g(\vec{x})^T
$$
As our transformation is component wise, the Jacobian matrix is diagonal too:
$$
(J_g)_{ij} = \frac{\partial g_i}{\partial y_j} = \begin{cases} g_i'(x_i) & \text{if } j=i \\ 0 & \text{if } j \neq i \end{cases}
$$
Therefore, $\mathbf{J}_g$ is a diagonal matrix, $\mathbf{J}_g^T = \mathbf{J}_g$, and the covariance matrix simplifies to
$$
\mathbf{\Sigma}_z \approx \mathbf{J}_g(\vec{x}) \mathbf{\Sigma}_y \mathbf{J}_g(\vec{x})
$$

\subsubsection*{Higher order moments and the small noise limit}
To justify our approximation we now show that its skewness and kurtosis are parametrically small. Because the transformation $z_i = g_i(y_i)$ is component wise, we can work on the marginal distribution of each component $z_i$ individually. The fluctuation $\delta z_i = z_i - \mu_{z,i}$ can be expanded up to third order as
$$
\delta z_i \approx {g_i'(x_i)\delta y_i}(\sigma_{y,i}) + \tfrac{1}{2}g_i''(x_i)(\delta y_i^2 - \sigma_{y,i}^2) + \tfrac{1}{6}g_i'''(x_i)(\delta y_i^3)
$$
We define the boundary distance $s_i \equiv \min\{x_i, \alpha_i - x_i\}$, noting that $g_i'(x_i) \sim \mathcal{O}(s_i^{-1})$ and $g_i''(x_i) \sim \mathcal{O}(s_i^{-2})$.

Skewness is the normalized third central moment. The dominant term for the variance is $\sigma_{z,i}^2 \approx (g_i')^2 \sigma_{y,i}^2$.
$$
\operatorname{Skewness}(z_i) = \frac{\mathbb{E}[\delta z_i^3]}{(\sigma_{z,i}^2)^{3/2}}
$$
We find the lowest order, nonzero term for the numerator is $3 (g_i'(x_i))^2 g_i''(x_i) \sigma_{y,i}^4$
Thus, the numerator scales as $\mathcal{O}(\sigma_{y,i}^4)$ which gives the total scaling of skewness with the boundary distance as:
$$
\operatorname{Skewness}(z_i) \sim \frac{\mathcal{O}(\sigma_{y,i}^4 / s_i^4)}{\mathcal{O}(\sigma_{y,i}^3 / s_i^3)} = \mathcal{O}\left(\frac{\sigma_{y,i}}{s_i}\right)
$$
Excess kurtosis is the normalized fourth cumulant, $\kappa_4$.
$$
\operatorname{Excess Kurtosis}(z_i) = \frac{\kappa_4(z_i)}{(\sigma_{z,i}^2)^2} = \frac{\mathbb{E}[\delta z_i^4] - 3(\mathbb{E}[\delta z_i^2])^2}{(\sigma_{z,i}^2)^2}
$$
The denominator scales as $(\mathcal{O}(\sigma_{y,i}^2))^2 = \mathcal{O}(\sigma_{y,i}^4)$. For the numerator, the dominant $\mathcal{O}(\sigma_{y,i}^4)$ terms cancel exactly. The numerator $\kappa_4(z_i)$ is dominated by the next highest order terms, which are $\mathcal{O}(\sigma_{y,i}^6)$. Including the boundary distance scaling is:
$$
\operatorname{Excess Kurtosis}(z_i) \sim \frac{\mathcal{O}(\sigma_{y,i}^6 / s_i^6)}{\mathcal{O}(\sigma_{y,i}^4 / s_i^4)} = \mathcal{O}\left(\frac{\sigma_{y,i}^2}{s_i^2}\right)
$$
    

Therefore, the distribution of $\vec{z} = \ln \vec{n}$ is approximately multivariate Gaussian in the small noise limit ($\sigma_{y,i} \ll s_i$) with the mean vector $\vec{\mu}_z$ and covariance matrix $\mathbf{\Sigma}_z$ given by
\begin{align}
    \mu_{z,i} &= g_i(x_i) + \frac{1}{2} g_i''(x_i) \Sigma_{y,ii} \\
    \mathbf{\Sigma}_z &= \mathbf{J}_g(\vec{x}) \mathbf{\Sigma}_y \mathbf{J}_g(\vec{x})
\end{align}
where $\mathbf{J}_g(\vec{x})$ is the diagonal matrix with entries $J_{ii} = g_i'(x_i)$.

By definition, the distribution of the equilibrium abundance vector $\vec{n} = e^{\vec{z}}$ is therefore approximately multivariate lognormal. Because marginal distribution of $\text{ln}(n_i)$ is also a Gaussian, we conclude that $n_i$ itself is approximately lognormal.

\bibliography{mainbib.bib}


\title{Feasibility as a moving target: Fluctuating species interactions lead to universal power law in equilibrium abundances \\
--Supplementary Material 2-- \\}
\author{Cagatay Eskin}
\email{ceskin@nd.edu}

\author{Vu Nguyen}
\email{}
\author{Dervis Can Vural}%
 \email{Corresponding author: dvural@nd.edu}
\affiliation{%
 Department of Physics and Astronomy, University of Notre Dame, Notre Dame, Indiana 46556, USA\\
}%

\date{\today}

\maketitle

\section{First Order Approximation to the Equilibrium}

We begin with the equilibrium condition for a generalized Lotka-Volterra system, where the population abundances $\vec{y}(t)$ are determined by the time-dependent interaction matrix $\mathbf{A}(t)$ and the intrinsic growth rates $\vec{r}$:
\begin{equation}
\label{eq:starting_equilibrium}
\mathbf{A}(t)\vec{y}(t)=-\vec{r} \quad \Rightarrow \quad \vec{y}(t)=-\mathbf{A}^{-1}(t)\vec{r}.
\end{equation}
Let the interaction matrix fluctuate around a constant mean $\mathbf{A}_{0}$ with a small perturbation $\delta \mathbf{A}(t)$:
\begin{equation}
\mathbf{A}(t)=\mathbf{A}_{0}+\delta \mathbf{A}(t).
\end{equation}
The equilibrium abundance without fluctuations is $\vec{x}=-\mathbf{A}_{0}^{-1}\vec{r}$. For small fluctuations $\delta \mathbf{A}(t)$, the inverse matrix $[\mathbf{A}_{0}+\delta \mathbf{A}(t)]^{-1}$ can be linearized using a geometric series expansion.
Noting that $(\mathbf{A}_{0}+\delta \mathbf{A})^{-1} = (\mathbf{I}+\mathbf{A}_{0}^{-1}\delta \mathbf{A})^{-1}\mathbf{A}_{0}^{-1}$, and using the approximation $(\mathbf{I}-\mathbf{X})^{-1} \approx \mathbf{I}+\mathbf{X} $ for small $\mathbf{X}$, we let $\mathbf{X} = -\mathbf{A}_{0}^{-1}\delta \mathbf{A}$. This yields the first-order approximation:
\begin{equation}
[\mathbf{A}_{0}+\delta \mathbf{A}(t)]^{-1}\approx{\mathbf{A}_{0}}^{-1}-{\mathbf{A}_{0}}^{-1}{\delta \mathbf{A}(t){\mathbf{A}_{0}}^{-1}}.
\end{equation}
The time-dependent equilibrium abundance $\vec{y}(t)$ is then:
\begin{align}
\vec{y}(t) &\approx -[{\mathbf{A}_{0}}^{-1}-{\mathbf{A}_{0}}^{-1}\delta \mathbf{A}(t) \mathbf{A}_{0}^{-1}]\vec{r} \notag \\
&= \vec{x} + \mathbf{A}_{0}^{-1}\delta \mathbf{A}(t) (\mathbf{A}_{0}^{-1}\vec{r}) \notag \\
&= \vec{x} - \mathbf{A}_{0}^{-1}\delta \mathbf{A}(t)\vec{x}.
\end{align}
The deviation from the mean equilibrium, $\delta\vec{y}(t) = \vec{y}(t)-\vec{x}$, is therefore:
\begin{equation}
\label{eq:deviation_vector}
\delta\vec{y}(t)\approx-\mathbf{A}_{0}^{-1}\delta \mathbf{A}(t)\vec{x}.
\end{equation}
In component form, defining $\mathbf{R} = \mathbf{A}_{0}^{-1}$, this is:
\begin{equation}
\delta y_{m}(t)\approx-\sum_{p,q=1}^{N}R_{mp}\delta A_{pq}(t)x_q.
\end{equation}
We can also define the sensitivity coefficients $S_{m,pq}$ as the partial derivative of the equilibrium abundance $y_m$ with respect to the matrix element $A_{pq}$, evaluated at $\mathbf{A}_0$. Using the matrix derivative rule $\partial \mathbf{A}^{-1}/\partial A_{pq} = -\mathbf{A}^{-1}(\partial \mathbf{A}/\partial A_{pq})\mathbf{A}^{-1} = -\mathbf{A}^{-1}\mathbf{E}_{pq}\mathbf{A}^{-1}$, where $\mathbf{E}_{pq}$ is an elementary matrix, we find:
\begin{equation}
S_{m,pq}=\left(\frac{\partial y_{m}}{\partial A_{pq}}\right)_{\mathbf{A}=\mathbf{A}_{0}} = -(\mathbf{A}_{0}^{-1}\mathbf{E}_{pq}\vec{x})_{m} = -R_{mp}x_{q}.
\end{equation}
The linearized deviation can then be written compactly as:
\begin{equation}
\label{eq:deviations_component_not}
\delta y_{m}(t) \approx \sum_{p,q=1}^{N}S_{m,pq}\delta A_{pq}(t).
\end{equation}

\section{Stochastic Differential Equation of the Equilibrium}

We model the fluctuations $\delta A_{pq}(t)$ as independent Ornstein-Uhlenbeck (OU) processes:
\begin{equation}
\label{eq:A_SDE}
d(\delta A_{pq}(t))=-k_{pq}\delta A_{pq}(t)dt+\sqrt{D_{pq}}dW_{pq}(t),
\end{equation}
where $k_{pq}$ is the relaxation rate, $D_{pq}$ is the noise strength, and $dW_{pq}(t)$ are independent Wiener processes. The stationary variance of this process is $\sigma_{A,pq}^{2}=D_{pq}/(2k_{pq})$.

Since each $\delta y_{m}$ is a linear sum of Gaussian processes, it is also a Gaussian process. We approximate its dynamics with a marginal OU process:
\begin{equation}
\label{eq:y_SDE}
d(\delta y_{k})=-\gamma_{k}\delta y_{k}dt+\sqrt{2D_{k,k}^{\text{eff}}}dW_{k}.
\end{equation}
Assuming the fluctuations $\delta A_{pq}$ are independent, the variance of $\delta y_k$ is:
\begin{equation}
\label{eq:var_y}
\sigma_{y,k}^{2} = \langle\delta y_k^2\rangle = \sum_{p,q=1}^{N}S_{k,pq}^{2} \langle \delta A_{pq}^2 \rangle = \sum_{p,q=1}^{N}S_{k,pq}^{2}\sigma_{A,pq}^{2}.
\end{equation}
This approximation neglects correlations between different $\delta y_k$ components.

The effective diffusion tensor $\mathbf{\mathcal{D}}^{\text{eff}}$ for the full vector process $\delta\vec{y}(t)$ is given by $(\mathbf{G}\mathbf{G}^{\top})/2$, where $\mathbf{G}$ is the noise matrix. From Eq.~\eqref{eq:deviations_component_not} and~\eqref{eq:A_SDE}, the noise term is $d(\delta y_{m})_{\text{noise}}=\sum_{p,q}S_{m,pq}\sqrt{D_{pq}}dW_{pq}(t)$. The elements of the diffusion tensor are:
\begin{equation}
\mathcal{D}_{mn}^{\text{eff}}=\frac{1}{2}\sum_{p,q=1}^{N} S_{m,pq}S_{n,pq}D_{pq}.
\end{equation}
We neglect the off-diagonal terms of this tensor and consider only the effective diffusion coefficients for each component:
\begin{equation}
D_{k,k}^{\text{eff}} = \frac{1}{2}\sum_{p,q=1}^{N}S_{k,pq}^{2}D_{pq}.
\end{equation}
Substituting $D_{pq}=2 k_{pq} \sigma_{A,pq}^{2}$, we get:
\begin{equation}
D_{k,k}^{\text{eff}}=\sum_{p,q=1}^{N}S_{k,pq}^{2}k_{pq}\sigma_{A,pq}^{2}.
\end{equation}
Finally, the effective relaxation rate $\gamma_k$ for the process $\delta y_k$ is found from the stationary variance relation $\sigma_{y,k}^2 = D_{k,k}^{\text{eff}}/\gamma_k$:
\begin{equation}
\label{eq:y_gamma}
\gamma_{k}=\frac{D_{k,k}^{\text{eff}}}{\sigma_{y,k}^{2}}=\frac{\sum_{p,q}S_{k,pq}^{2}k_{pq}\sigma_{A,pq}^{2}}{\sum_{p,q}S_{k,pq}^{2}\sigma_{A,pq}^{2}}.
\end{equation}
This shows that $\gamma_k$ is a weighted average of the underlying relaxation rates $k_{pq}$.

\section{Fokker-Planck Equation and Mean Exit Time}

The SDE in Eq.~\eqref{eq:y_SDE} corresponds to the Fokker-Planck equation for the probability density $p(\delta y_{k},t)$:
\begin{equation}
\label{eq:FPE_y}
\frac{\partial p}{\partial t} = \frac{\partial}{\partial\delta y_{k}}(\gamma_{k}\delta y_{k}p) + D_{k,k}^{\text{eff}}\frac{\partial^{2}p}{\partial\delta y_{k}^{2}}.
\end{equation}
This is an equation of the form $\partial_t p = -\partial_x [A(x)p] + \frac{1}{2}\partial_x^2 [B(x)p]$, with drift $A(x) = -\gamma_k x$ and diffusion $B(x) = 2D_{k,k}^{\text{eff}}$.

We are interested in the mean time $T_k(x_0)$ for a particle starting at $x_0 = \delta y_k(0)$ to first reach an absorbing boundary. This is governed by the Pontryagin (or backward Fokker-Planck) equation:
\begin{equation}
A(x)\frac{dT}{dx}+\frac{1}{2}B(x)\frac{d^{2}T}{dx^{2}}=-1.
\end{equation}
For a process on an interval $[b, a]$ with an absorbing boundary at $b$ and a reflecting boundary at $a$, the solution for the mean exit time from a starting point $x$ is given by Gardiner as:
\begin{equation}
T(x)=2\int_{b}^{x}\frac{dy}{\Psi(y)}\int_{y}^{a}\frac{\Psi(z)}{B(z)}dz,
\end{equation}
where $\Psi(x)=\exp\left[ \int \frac{2A(x')}{B(x')} dx' \right]$.

In our case, the process for $\delta y_k$ starts at 0. Feasibility is lost if any abundance $y_k = x_k + \delta y_k$ hits zero, so the absorbing boundary is at $\delta y_k = -x_k$. We set the reflecting boundary at infinity. The potential function $\Psi$ is:
\begin{equation}
\Psi(\delta y_{k})=\exp\left(\int_{0}^{\delta y_k}\frac{-\gamma_k z}{D_{k,k}^{\text{eff}}}dz\right) = \exp\left(-\frac{\gamma_k (\delta y_k)^2}{2D_{k,k}^{\text{eff}}}\right).
\end{equation}
Using $\gamma_k = D_{k,k}^{\text{eff}}/\sigma_{y,k}^2$, this simplifies to:
\begin{equation}
\label{eq:y_potential}
\Psi(\delta y_{k})=\exp\left(-\frac{(\delta y_{k})^{2}}{2\sigma_{y,k}^{2}}\right).
\end{equation}
We want the mean exit time starting from $\delta y_k(0)=0$ to the absorbing boundary at $b=-x_k$, with a reflecting boundary at $a=\infty$. The formula gives:
\begin{equation}
\label{eq:y_MTE_double_int}
T_{k}(0) = \frac{1}{D_{k,k}^{\text{eff}}}\int_{-x_{k}}^{0}e^{\frac{z_{1}^{2}}{2\sigma_{y,k}^{2}}} dz_{1}\int_{z_{1}}^{\infty}e^{-\frac{z_{2}^{2}}{2\sigma_{y,k}^{2}}}dz_{2}.
\end{equation}
The inner integral can be expressed using the complementary error function, $\text{erfc}(u) = (2/\sqrt{\pi})\int_u^\infty e^{-t^2}dt$:
\begin{equation*}
\int_{z_{1}}^{\infty}e^{-\frac{z_{2}^{2}}{2\sigma_{y,k}^{2}}}dz_{2} = \frac{\sqrt{2\pi}\sigma_{y,k}}{2}\text{erfc}\left(\frac{z_{1}}{\sqrt{2}\sigma_{y,k}}\right).
\end{equation*}
Thus, the mean exit time is given by the single integral:
\begin{equation}
\label{eq:y_MTE_single_integral}
T_{k}(0)=\frac{\sqrt{2\pi}\sigma_{y,k}}{2D_{k,k}^{\text{eff}}}\int_{-x_{k}}^{0}e^{\frac{z_{1}^{2}}{2\sigma_{y,k}^{2}}}\text{erfc}\left(\frac{z_{1}}{\sqrt{2}\sigma_{y,k}}\right)dz_{1}.
\end{equation}

\section{Asymptotic Approximation of Exit Time}

We evaluate the integral in Eq.~\eqref{eq:y_MTE_single_integral} using Laplace's method for large barriers, i.e., when $\Lambda \equiv x_k^2 / (2\sigma_{y,k}^2) \gg 1$. The integral is of the form $\int_a^b g(t)e^{\lambda f(t)}dt$. Here, $\lambda = 1/(2\sigma_{y,k}^2)$, $f(z_1) = z_1^2$, and $g(z_1) = \text{erfc}(z_1/(\sqrt{2}\sigma_{y,k}))$.

The maximum of $f(z_1)$ on the interval $[-x_k, 0]$ occurs at $z_1 = -x_k$. We approximate the integral by expanding the exponent around this maximum. Let $z_1 = -x_k + \xi$:
\begin{equation}
I \approx \int_{0}^{x_k} g(-x_k+\xi) e^{\lambda(-x_k+\xi)^2} d\xi.
\end{equation}
For large $\lambda$, the main contribution comes from $\xi \approx 0$. We approximate $g(-x_k+\xi) \approx g(-x_k)$ and $(-x_k+\xi)^2 \approx x_k^2 - 2x_k\xi$.
\begin{align}
I &\approx g(-x_k) e^{\lambda x_k^2} \int_0^{x_k} e^{-2\lambda x_k \xi} d\xi \notag \\
&\approx g(-x_k) e^{\lambda x_k^2} \int_0^\infty e^{-2\lambda x_k \xi} d\xi \notag \\
&= g(-x_k) e^{\lambda x_k^2} \frac{1}{2\lambda x_k}.
\end{align}
The slowly varying part is $g(-x_k) = \text{erfc}(-x_k/(\sqrt{2}\sigma_{y,k}))$. For a large positive argument $Z = x_k/(\sqrt{2}\sigma_{y,k})$, $\text{erfc}(-Z) = 2-\text{erfc}(Z) \approx 2$.
Plugging this back into the expression for $T_k(0)$:
\begin{align}
T_{k}(0) &\approx \frac{\sqrt{2\pi}\sigma_{y,k}}{2D_{k,k}^{\text{eff}}} \left( 2   e^{\lambda x_k^2}  \frac{1}{2\lambda x_k} \right) \notag \\
&= \frac{\sqrt{2\pi}\sigma_{y,k}}{D_{k,k}^{\text{eff}}} \frac{e^{x_k^2/(2\sigma_{y,k}^2)}}{ (1/\sigma_{y,k}^2) x_k }
\end{align}
Using $D_{k,k}^{\text{eff}} = \gamma_k \sigma_{y,k}^2$, we arrive at the final approximation:
\begin{equation}
T_{k}(0) \approx \frac{\sigma_{y,k}\sqrt{2\pi}}{\gamma_k x_k} \exp\left(\frac{x_k^2}{2\sigma_{y,k}^2}\right).
\end{equation}

\section{Escape Rates and Mean Time to Feasibility Loss}

The escape rate $\lambda_k$ for the $k$-th species is the inverse of its mean exit time:
\begin{equation}
\label{eq:escape_rates}
\lambda_{k} = \frac{1}{T_{k}(0)} \approx \frac{\gamma_{k} x_{k}}{\sigma_{y,k}\sqrt{2\pi}}\exp\left(-\frac{x_{k}^{2}}{2\sigma_{y,k}^{2}}\right).
\end{equation}
Treating the extinction of each species as an independent Poisson process, the total failure rate for the system is the sum of the individual rates, $\lambda_{\text{eff}}=\sum_{k} \lambda_{k}$. The mean time to feasibility loss for the entire ecosystem is the inverse of this total rate:
\begin{equation}
\label{eq:T_loss}
\langle T_{\text{loss}} \rangle \approx \frac{1}{\lambda_{\text{eff}}} = \left(\sum_{k=1}^{N}\frac{\gamma_{k} x_{k}}{\sigma_{y,k}\sqrt{2\pi}}\exp\left(-\frac{x_{k}^{2}}{2\sigma_{y,k}^{2}}\right)\right)^{-1}.
\end{equation}